\documentclass[12pt,a4paper]{iopart}
\usepackage[english]{babel}

\usepackage{floatrow}
\usepackage{iopams}
\usepackage{verbatim}
\usepackage[usenames,dvipsnames]{xcolor}
\usepackage{graphicx}
\usepackage{float}
\usepackage{color}
\usepackage[colorlinks,citecolor=blue,linkcolor=blue,urlcolor=blue]{hyperref}
\usepackage{amssymb}
\usepackage{hyperref}
\usepackage[utf8]{inputenc}

\def\iotabar{\lower3pt\hbox{$\mathchar'26$}\mkern-7mu\iota}

\newcounter{appnumb}

\newcommand{{\casdk}}{{\small CAS3D-K}}
\newcommand{{\eut}}{{\small EUTERPE}}
\newcommand{{\gene}}{{\small GENE}}

\begin{document}

\title[Semianalytical calculation of the zonal-flow oscillation frequency in stellarators]{Semianalytical calculation of the zonal-flow oscillation frequency in stellarators}
 \vspace{-0.3cm}
\author{Pedro Monreal$^{1}$} \vspace{-0.2cm} 
\eads{\mailto{pedro.monreal@ciemat.es}} \vspace{-0.5cm} 
\author{Edilberto S\'anchez$^{1}$} \vspace{-0.2cm}
\eads{\mailto{edi.sanchez@ciemat.es}} \vspace{-0.5cm} 
\author{Iv\'an Calvo$^{1}$} \vspace{-0.2cm}
\eads{\mailto{ivan.calvo@ciemat.es}} \vspace{-0.5cm}
\author{Andr\'es Bustos$^{2}$} \vspace{-0.2cm}
\eads{\mailto{anbustos@fis.uc3m.es}} \vspace{-0.5cm}
\author{F\'elix I Parra$^{3,4}$} \vspace{-0.2cm}
 \eads{\mailto{felix.parradiaz@physics.ox.ac.uk}} \vspace{-0.5cm} 
\author{Alexey Mishchenko$^{5}$} \vspace{-0.2cm}
\eads{\mailto{alexey.mishchenko@ipp.mpg.de}} \vspace{-0.5cm} 
\author{Axel K\"onies$^{5}$} \vspace{-0.2cm}
\eads{\mailto{axel.koenies@ipp.mpg.de}} \vspace{-0.5cm} 
\author{Ralf Kleiber$^{5}$} \vspace{-0.2cm}
\eads{\mailto{ralf.kleiber@ipp.mpg.de}} \vspace{-0.5cm} 

\vspace{0.75cm}

\address{$^1$Laboratorio Nacional de Fusi\'on, CIEMAT, 28040 Madrid, Spain}
\address{$^2$Departamento de F\'isica, Universidad Carlos III de Madrid, 28911 Legan\'es, Spain}
\address{$^3$Rudolf Peierls Centre for Theoretical Physics, University of Oxford, Oxford, OX1 3NP, UK}
\address{$^4$Culham Centre for Fusion Energy, Abingdon, OX14 3DB, UK}
\address{$^5$Max-Planck-Institut f\"ur Plasmaphysik, D-17491 Greifswald, Germany}

\vskip 0.25cm


\vspace{-0.5cm}

\begin{abstract}
Due to their capability to reduce turbulent transport in magnetized plasmas, understanding the dynamics of zonal flows is an important problem in the fusion programme. Since the pioneering work by Rosenbluth and Hinton in axisymmetric tokamaks, it is known that studying the linear and collisionless relaxation of zonal flow perturbations gives valuable information and physical insight. Recently, the problem has been investigated in stellarators and it has been found that in these devices the relaxation process exhibits a characteristic feature: a damped oscillation. The frequency of this oscillation might be a relevant parameter in the regulation of turbulent transport, and therefore its efficient and accurate calculation is important.  Although an analytical expression can be derived for the frequency, its numerical evaluation is not simple and has not been exploited systematically so far. Here, a numerical method for its evaluation is considered, and the results are compared with those obtained by calculating the frequency from gyrokinetic simulations. This ``semianalytical" approach for the determination of the zonal-flow frequency reveals accurate and faster than the one based on gyrokinetic simulations.
\end{abstract}

\maketitle

\section{Introduction}
\label{sec:Introduction}

The reduction of turbulent transport by shearing of zonal flows in tokamaks and stellarators has been studied for decades now \cite{Hasegawa79, Diamond2005}, and their presence has been observed experimentally in several fusion devices \cite{Fujisawa2009}. These flows are associated to electrostatic potential perturbations constant on flux surfaces, which are generated from drift-waves by inverse cascade of energy in a turbulent plasma. Although the generation of zonal flows by plasma turbulence is a non-linear process, the study of its linear relaxation provides insight into the problem at an affordable cost. Furthermore, the linear evolution of zonal flows could be relevant for the saturation and regulation of the turbulence in some situations; e.g. under marginal stability conditions in which the turbulence drive, and consequently the zonal-flow drive, is small.

The linear evolution of zonal flows is greatly influenced by the magnetic geometry. Therefore, understanding its behavior in different geometries could help in the design of future devices optimized for enhanced zonal flows and reduced turbulent transport. Rosenbluth and Hinton \cite{Rosenbluth1998} studied the evolution of an initial zonal potential perturbation with small wavenumber in tokamak geometry, showing that the perturbation is not completely damped by collisionless processes but reaches a finite value at long times, the so-called zonal-flow residual level. Since then, the long-time collisionless evolution of zonal potential perturbations\footnote{That is, electrostatic potential perturbations that are constant on flux surfaces, while having a finite radial scale.} has attracted the attention of different authors that continued the analysis of the problem in axisymmetric configurations \cite{Xiao2006, Xiao2007, Jenko2000} and more recently in stellarators \cite{Sugama2005,Sugama2006,Sugama2007,Mishchenko2008,Helander2011,Xanthopoulos2011,Monreal2016}. In \cite{Monreal2016} a method for the fast numerical calculation of the residual level in tokamaks and stellarators for arbitrary wavenumbers of the perturbation was presented.

In general, the collisionless evolution of an initial zonal-flow perturbation in a stellarator involves two oscillations with different typical values of the frequency and also different physical origin.  On the one hand, a decaying geodesic acoustic mode (GAM) oscillation \cite{Winsor1968}, mainly caused by the dynamics of passing ions (at least in large aspect ratio devices). The collisionless damping of the GAM strongly depends on the safety factor \cite{Sugama2005,Sugama2007,Gao2008} and the frequency, $\Omega_{\rm GAM}$, is on the order of the characteristic frequencies of the turbulence; that is, $\Omega_{\rm GAM} \sim v_{ti}/L$. Here, $v_{ti}$ is the ion thermal speed and $L$ is a characteristic macroscopic scale of the system\footnote{Typically, $L\sim R$, the major radius of the torus.}. On the other hand, in \cite{Mishchenko2008} it was found that in stellarator geometry an additional oscillation takes place. Its frequency, that we denote by $\Omega_{\rm ZF}$, is significantly smaller than $\Omega_{\rm GAM}$. Unlike the GAM oscillation, that happens in tokamaks and stellarators, this slower oscillation is characteristic of stellarator geometries\footnote{It is more precise to say that the oscillation with frequency $\Omega_{\rm ZF}$ exists only in non-omnigeneous devices. In particular, it can also manifest in rippled tokamaks. This will be explained below.} and has been experimentally observed in \cite{Alonso16}. If we denote by $\mathbf{v}_{d i}$ the ion magnetic drift and by $\psi$ a radial coordinate,
\begin{equation}\label{eq:OmegaZFtypicalvalue}
  \Omega_{\rm ZF} \sim 
  \frac{\overline{\mathbf{v}_{d i}\cdot\nabla\psi}}{\mathbf{v}_{d i}\cdot\nabla\psi} \, \frac{v_{ti}}{L}.
\end{equation}
Here, $\overline{\mathbf{v}_{d i}\cdot\nabla\psi}$ stands for a typical value of the radial magnetic drift averaged over a trapped trajectory. Since, in general, $\overline{\mathbf{v}_{d i}\cdot\nabla\psi} / \mathbf{v}_{d i}\cdot\nabla\psi \ll 1$, we have $\Omega_{\rm ZF}\ll \Omega_{\rm GAM}$. From now on, when we refer to a zonal flow oscillation (or simply to an oscillation), and if not stated otherwise, we will understand that we are talking about the low frequency one, which is the main subject of this paper.

The interest in the computation of $\Omega_{\rm ZF}$ resides in the potential role of the oscillation for the regulation of turbulent transport, pointed out in \cite{Xanthopoulos2011}. Since the residual value of the zonal flow is reached at times longer than the typical saturation time of the turbulence, the oscillatory phase of the zonal flow relaxation is likely to be physically more important than the value achieved when $t\to \infty$.

An analytical expression for $\Omega_{\rm ZF}$ was first derived in \cite{Mishchenko2008}. The oscillation was further studied, including its damping, in \cite{Helander2011}. As we will see in Section~\ref{sec:Evolution}, the derivation of the expression for $\Omega_{\rm ZF}$ employs a local gyrokinetic equation. The expansions used to arrive at the expression for $\Omega_{\rm ZF}$ require
\begin{equation}\label{eq:validityapproximations}
  \frac{1}{L} \ll k_\perp \ll \frac{1}{\rho_{t i}},
\end{equation}
where $k_\perp$ is the wavenumber of the zonal potential perturbation, $\rho_{t i} = v_{ti} / \Omega_i$ is the thermal ion gyroradius, $\Omega_i = Z_i e B / m_i$ is the ion gyrofrequency, $Z_i e$ is the ion charge, $e$ is the proton charge, $B$ is the magnitude of the magnetic field $\mathbf{B}$ and $m_i$ is the ion mass. In reference \cite{Helander2011}, it was pointed out that it is unclear whether or not the analytical expression for the frequency is quantitatively accurate in actual devices. Anyway, the accuracy of the analytical expression has not been systematically checked so far, mainly because its evaluation is non-trivial: it involves phase-space averages that cannot be computed analytically in stellarator geometry, and therefore the evaluation must be carried out numerically. Here, we investigate this ``semianalytical" method to determine the value of $\Omega_{\rm ZF}$ by using an extension \cite{Monreal2016} of {\casdk} \cite{koenies2000,koenies2008}. The results of the semianalytical calculation are compared to those obtained from gyrokinetic simulations with the radially global code {\eut} \cite{Jost2001,Kleiber2012} and the radially local code {\gene} \cite{Jenko2000,Gorler2011,GENE,Xanthopoulos2009} in the W7-X, TJ-II and LHD stellarators and also in a series of rippled tokamak equilibria.

The paper is organized as follows. In Section~\ref{sec:Evolution} we derive the expression for the frequency of the zonal flow oscillation, $\Omega_{\rm ZF}$. In Section~\ref{sec:Calculation}, we describe the numerical methods to evaluate this expression with the code {\casdk} and we comment on how to obtain $\Omega_{\rm ZF}$ from gyrokinetic simulations with the global gyrokinetic code {\eut} and the full flux-surface version of {\gene}. In Section \ref{sec:Tokamakresults} we compute $\Omega_{\rm ZF}$, with both methods (that is, the semianalytical approach and the one based on direct gyrokinetic simulations), in a tokamak configuration with different ripple values, which we use as a test to gain insight into the problem and check the limitations of the numerical methods. As explained in Section~\ref{sec:Tokamakresults}, we will use an `academic' tokamak configuration, with unrealistic safety factor profile, so that in the simulations the zonal flow oscillation is neatly observed and we can clearly illustrate how $\Omega_{\rm ZF}$ varies as a function of the ripple size. In Section~\ref{sec:Stellaratorresults} we calculate the zonal-flow frequency in the W7-X, TJ-II and LHD stellarators. The computational requirements of the two approaches to the calculation of $\Omega_{\rm ZF}$ are compared in Section~\ref{sec:simulations}. Finally, the conclusions are given in Section~\ref{sec:conclusions}.

\section{Derivation of the analytical expression for the zonal-flow oscillation frequency}
\label{sec:Evolution}

In this section we briefly explain how to calculate the long-time, linear and collisionless evolution of a zonal potential perturbation. More details can be found in \cite{Monreal2016}. The focus here will be on the derivation of an expression for $\Omega_{\rm ZF}$ and on the approximations made to obtain it. In later sections the numerical evaluation of this expression is compared to gyrokinetic simulations for several toroidal devices.

\subsection{Long-time collisionless relaxation of zonal electrostatic perturbations}
\label{sec:longtimeevolutionZF}

We use straight-field-line coordinates $\{\psi,\theta,\zeta\}$ to locate a point in space, $\mathbf{x}(\psi,\theta,\zeta)$. Here, $\psi=\Psi_t/\Psi_t^\mathrm{edge}$ is the radial coordinate, defined as the toroidal flux $\Psi_t$ normalized by its value at the last closed flux surface $\Psi_t^\mathrm{edge}$, and $\theta$ and $\zeta$ are, respectively, poloidal and toroidal angles normalized such that $\theta,\zeta\in [0,1)$. In these coordinates, the contravariant form of the magnetic field reads
\begin{equation} \label{eq:Bcontravariantform}
  {\bf B} = 
  -\Psi_p'(\psi)\nabla \psi \times \nabla (\zeta - q(\psi)\theta),
\end{equation}
where $q(\psi)=\Psi'_t(\psi)/\Psi'_p(\psi)$ is the safety factor and $\Psi_p'(\psi)$ and $\Psi_t'(\psi)$ are the derivatives of the poloidal and toroidal fluxes with respect to $\psi$. It is convenient to define the coordinate $\alpha = \zeta - q(\psi) \theta$, that labels magnetic field lines on a surface once $\psi$ has been fixed. Unless otherwise specified, below we use  $\{\psi,\theta,\alpha\}$ as independent spatial coordinates. Observe that, given $\psi$ and $\alpha$, the coordinate $\theta$ locates the point along the field line. The independent velocity coordinates (all functions in this paper are independent of the gyrophase) are $\{v,\lambda,\sigma\}$, where $v$ is the magnitude of the velocity, $\lambda=B^{-1}v_\perp^2/v^2$ is the pitch-angle coordinate, $v_\perp$ is the component of the velocity perpendicular to the magnetic field and $\sigma=v_\parallel/|v_\parallel|$ is the sign of the parallel velocity,
\begin{equation}
  v_\parallel(\psi,\theta,\alpha,v,\lambda,\sigma) = 
  \sigma v \sqrt{1-\lambda B(\psi,\theta,\alpha)}\, .
\end{equation}

We are interested in studying the linear and collisionless evolution of an electrostatic potential perturbation of the form
\begin{equation}\label{eq:eikonalvarphi}
  \varphi = \varphi_k(\psi,t) \mathrm{e}^{\mathrm{i}k_\psi \psi}.
\end{equation}
Let us denote the phase-space distribution function of species $s$ by $F_s=F_s(\mathbf{x},v,\lambda,\sigma,t)$ and the thermal gyroradius by $\rho_{ts} = v_{ts} / \Omega_s$, where $v_{ts}=\sqrt{T_s/m_s}$, $T_s$ and $m_s$ are the thermal speed, temperature and mass of species $s$, and $\Omega_s = Z_s e B / m_s$ and $Z_s e$ are the gyrofrequency and charge of species $s$. We expand $F_s$ in  $\rho_{ts\star} = \rho_{ts} / L \ll 1$,
\begin{equation}
  F_s(\mathbf{x},v,\lambda,\sigma,t) = 
  F_{s0}(\mathbf{x},v) + 
  F_{s1}(\mathbf{x},v,\lambda,\sigma,t) + 
  O(\rho_{ts\star}^2F_{s0}),
\end{equation}
where $F_{s1} \sim O(\rho_{ts\star}F_{s0})$ and $F_{s0}$ is a Maxwellian distribution with density $n_s(\psi)$ and temperature $T_s(\psi)$ constant on flux surfaces. That is,
\begin{equation}
  F_{s0}(\mathbf{x},v) := 
  \frac{n_s(\psi(\mathbf{x}))}{(\sqrt{2\pi} v_{ts}(\psi(\mathbf{x})))^3}
  \exp\left(-\, \frac{v^2}{2v^2_{ts}(\psi(\mathbf{x}))}\right),
\end{equation}
with $\sum_s Z_s n_s = 0$ due to quasineutrality. For the sake of consistence with (\ref{eq:eikonalvarphi}), we also assume that the perturbation to the  Maxwellian can be expressed as
\begin{equation}\label{eq:eikonalF1}
  F_{s1}(\mathbf{x},v,\lambda,\sigma,t) = f_s(\mathbf{x},v,\lambda,\sigma,t) \mathrm{e}^{\mathrm{i}k_\psi \psi(\mathbf{x})}.
\end{equation}
Then, the collisionless gyrokinetic equation, written in terms of the non-adiabatic component
\begin{equation}
  h_s = f_s + \frac{Z_s e}{T_s} \varphi_k J_{0}(k_\perp\rho_s)  F_{s0},
\end{equation}
reads \cite{Monreal2016} 
\begin{eqnarray}
  \left(\partial_t + v_\parallel\, \hat\mathbf{b} \cdot \nabla 
  + \mathrm{i}k_\psi \omega_s\right) h_s 
  = \frac{Z_s e}{T_s} J_{0}(k_\perp\rho_s) F_{s0} \partial_t \varphi_k.
  \label{eq:gk}
\end{eqnarray}
Here, $k_\perp:=k_\psi|\nabla\psi|$, $\hat\mathbf{b}$ is the unit vector along the magnetic field, $J_0$ is the zeroth-order Bessel function of the first kind and $\rho_s=v_\perp/\Omega_s$ is the gyroradius of species $s$.  Below, we often use the short-hand notation $J_{0s}\equiv J_0(k_\perp\rho_s)$. Finally, $\omega_s := \mathbf{v}_{ds}\cdot\nabla \psi$ is the radial magnetic drift frequency of species $s$, where
\begin{equation}
  \mathbf{v}_{ds} =
  \frac{v^2}{\Omega_s} \hat\mathbf{b} \times 
  \left[(1-\lambda B)\hat\mathbf{b}\cdot\nabla
    \hat\mathbf{b} + 
    \frac{\lambda}{2}
    \nabla B \right]
  \label{eq:driftv}
\end{equation}
is the magnetic drift velocity.

Equation (\ref{eq:gk}) is solved along with the flux-surface averaged quasineutrality equation
\begin{equation}
  \sum_s \frac{Z_s^2 e}{T_s}n_s\, {\varphi}_k = 
  \left\langle 
  \sum_s Z_s \int J_{0s} {h}_s
  \mathrm{d}^3 v  
  \right\rangle_\psi.
  \label{eq:qn2}
\end{equation}
The flux-surface average is defined, for any given function $G=G(\psi,\theta,\zeta)$, as
\begin{equation}
  \langle G \rangle_\psi = 
  V'(\psi)^{-1} \int_0^{1} \mathrm{d}\theta \int_0^{1} 
  \mathrm{d} \zeta \sqrt{g}\, G(\psi,\theta,\zeta),
  \label{eq:fsa}
\end{equation}
where $\sqrt{g} = [\nabla\psi\cdot(\nabla\theta\times\nabla\zeta)]^{-1}$ is the square root of the metric determinant and $V'(\psi) = \int_0^{1} \mathrm{d} \theta \int_0^{1} \mathrm{d} \zeta \sqrt{g}$ is the derivative of the volume enclosed by the flux surface labeled by $\psi$.

Note that equation (\ref{eq:gk}) is based on an eikonal representation of the fields (see (\ref{eq:eikonalvarphi}) and (\ref{eq:eikonalF1})), where the variation on small scales is contained in $\exp(\mathrm{i} k_\psi \psi)$, and $\varphi_k$, $f_s$ and $h_s$ vary on the macroscopic scale $L$. Equation (\ref{eq:gk}) is valid as long as \cite{Monreal2016}
\begin{equation}\label{eq:conditions_validity_gkeq}
  \frac{1}{L}\ll k_\perp \lesssim \frac{1}{\rho_{ts}}.
\end{equation} 

Before proceeding to give the solution of equations (\ref{eq:gk}) and (\ref{eq:qn2}) at long times, it seems timely to discuss why the role of the background radial electric field, $E_\psi$, has not been considered. The background radial electric field enters the collisionless gyrokinetic equation via terms like $\mathbf{v}_E\cdot\nabla\alpha\partial_\alpha h_s$, where $\mathbf{v}_E\cdot\nabla\alpha = E_\psi/\Psi'_p$ is the component of the $E\times B$ drift $\mathbf{v}_E$ in the direction $\alpha$. This term is smaller than the last term on the left side of (\ref{eq:gk}) as long as
\begin{equation}
\frac{L \mathbf{v}_E\cdot\nabla\alpha}{v_{ts}} < k_\perp \rho_{ts}.
\end{equation}
Hence, when $k_\perp \rho_{ts}\sim 1$ the effect of  $E_\psi$ can be safely neglected\footnote{This is  true in the strictly collisionless limit and for the time scales considered in this paper. It is known that the background radial electric field is essential to calculate the neoclassical equilibrium in the $\sqrt{\nu}$ collisionality regime, for example~\cite{Calvo2017}.}. Of course, when $k_\perp$ approaches the value $1/L$ (but recall that (\ref{eq:conditions_validity_gkeq}) has to be satisfied), the background radial electric field can introduce corrections to the results that we will present in this article. However, a rigorous treatment of these corrections exceeds the scope of this work. In particular, if $E_\psi$ is included, the partial differential equation to be solved has higher dimensionality because differential terms in $\alpha$ appear. Some analytical progress in this direction is provided in \cite{Mishchenko2012}, using a model for stellarator geometry and making some additional simplifying assumptions. In what follows, we restrict ourselves to the set of equations consisting of (\ref{eq:gk}) and (\ref{eq:qn2}).

In order to solve equations (\ref{eq:gk}) and (\ref{eq:qn2}) it is convenient to use the Laplace transform, defined for any function $Q(t)$ as $\widehat{Q}(p) = \int_0^\infty Q(t) \mathrm{e}^{-pt} \mathrm{d}t$, where $p$ denotes the variable in Laplace space.  Solving for $p / (v_{ts}L^{-1}) \ll 1$, we find that \cite{Monreal2016}
\begin{eqnarray}
  \widehat{\varphi}_k(p) = 
  \frac{
    \sum_s Z_s\left\{ \frac{1}{p+\mathrm{i}k_\psi\overline{\omega_s}}
    \mathrm{e}^{-\mathrm{i}k_\psi\delta_s} J_{0s} 
    \overline{\mathrm{e}^{\mathrm{i}k_\psi\delta_s} f_s(0) } /F_{s0}
    \right\}_s
  }{
    \sum_s \frac{Z_s^2 e}{T_s} 
    \left\{1 - \frac{p}{p+\mathrm{i}k_\psi\overline{\omega_s}} \,
    \mathrm{e}^{-\mathrm{i}k_\psi\delta_s} J_{0s} 
    \overline{ \mathrm{e}^{\mathrm{i}k_\psi\delta_s} J_{0s}} \right\}_s  
  },
  \label{eq:gk6}
\end{eqnarray}
where $f_s(0)\equiv f_s(\mathbf{x},0)$ is the initial condition for $f_s$ and we have simplified the notation by defining the phase-space integration
\begin{equation}
  \left\{Q\right\}_s  := 
  \left\langle 
  \sum_{\sigma=-1}^1 \int_0^\infty \mathrm{d}v\,\int_0^{1/B} 
  \mathrm{d}\lambda\, \frac{\pi v^2 B}{\sqrt{1-\lambda B}}
  Q (\psi,\theta,\alpha,v,\lambda,\sigma)\,
  F_{s0} \right\rangle_\psi
  \label{eq:corchete}
\end{equation}
for any phase-space function $Q$.

The derivation of (\ref{eq:gk6}), and the expression (\ref{eq:gk6}) itself, makes use of  the orbit average, defined for a function $Q(\psi,\theta,\alpha,v,\lambda,\sigma,t)$ by
\begin{equation}
  \overline{Q} :=
  \left\{
  \begin{array}{lcr}
    \langle B\, Q /|v_\parallel |\rangle_\psi /
    \langle B /|v_\parallel|\rangle_\psi &&
    \textrm{passing particles,\,\,}\vspace{0.2cm} \\
    \omega_b\oint \mathrm{d}\theta \,Q /(v_\parallel\,
    \hat\mathbf{b} \cdot \nabla \theta) && 
    \textrm{trapped particles,}
  \end{array}
  \right.
  \label{eq:bounceaverages}
\end{equation}
where $\omega_b := [\oint \mathrm{d}\theta / (v_\parallel\, \hat\mathbf{b}\cdot\nabla \theta)]^{-1}$ is the bounce frequency. The orbit average removes frequencies that are $O(v_{ts} / L)$ or higher, and in particular it removes the GAM oscillation. Here, the symbol $\oint$ stands for integration (at fixed $\psi$ and $\alpha$) over the trapped trajectory.

The quantity $\delta_s$ entering (\ref{eq:gk6}) is defined as follows.
The orbit average acting on the parallel streaming operator has the useful property
\begin{equation}
  \overline{v_\parallel \hat\mathbf{b}\cdot\nabla Q} = 0
\end{equation}
for any single-valued function $Q$. Using this property, we can write the radial magnetic drift frequency as
\begin{equation}
  \omega_s = 
  \overline{\omega_s}
  +v_\parallel\, \hat\mathbf{b}\cdot\nabla \delta_s,
  \label{eq:mde}
\end{equation}
where $\delta_s = \delta_s(\psi,\theta,\alpha,v,\lambda,\sigma)$, that we choose to be odd in $\sigma$, is the radial displacement of the particle's gyrocenter from its mean flux surface.

Note that in axisymmetric tokamaks and in omnigenous stellarators $\overline{\omega_s}=0$ holds for all trajectories; in a tokamak with ripple and in non-omnigenous stellarators, one can only guarantee $\overline{\omega_s}= 0$ for passing particles. In subsection \ref{sec:delta_s} we explain how to calculate $\delta_s$ from (\ref{eq:mde}).

\subsection{Calculation of $\delta_s$}
\label{sec:delta_s}

In this work, we have employed and implemented in CAS3D-K a faster method to compute $\delta_s$ for trapped trajectories with respect to that used in \cite{Monreal2016}. We point out this below.

If we take $\theta$ and $\zeta$ to be Boozer angles, the contravariant form of $\mathbf{B}$ is given by (\ref{eq:Bcontravariantform}) and its covariant form is given by
\begin{equation}
  {\bf B}
  = I_t \nabla \theta 
  - I_p \nabla\zeta
  + \widetilde{\beta} \left(\psi,\theta,\zeta\right)\nabla\psi,
\end{equation}
with $I_t=I_t(\psi)$ and $I_p=I_p(\psi)$ the toroidal and poloidal currents, respectively. It is straightforward to realize that the metric determinant $\sqrt{g} = [\nabla\psi\cdot(\nabla\theta\times\nabla\zeta)]^{-1}$ can be written as 
\begin{equation}
  \sqrt{g} =  \frac{I_t\Psi_p' - I_p \Psi_t'}{B^2}.
\end{equation}

We split the solution to (\ref{eq:mde}) as
\begin{equation}
  \delta_s =
  -\frac{I_p}{\Psi_p'}\rho_{\parallel s} 
  +\widetilde{\delta_s}.
  \label{eq:mdet3}
\end{equation}
Following \cite{Monreal2016}, we have that for passing particles
\begin{equation}
  \widetilde{\delta_s}
  = \left(\frac{I_p}{\Psi_p'}-\frac{I_t}{\Psi_t'}\right) \sum_{m,n\neq0}
  \left( \frac{qn}{m + qn} \right)
       {(\rho_{\parallel s})}_{mn} \mathrm{e}^{ 2\pi
         \mathrm{i}(m\theta+n\zeta)},
       \label{eq:deltap}
\end{equation}
where
\begin{equation}
  \rho_{\parallel s} = \sum_{m,n} 
  (\rho_{\parallel s})_{mn}
  \mathrm{e}^{2\pi \mathrm{i}(m\theta+n\zeta)}.
\end{equation}

For trapped particles
\begin{equation}\label{eq:delta_s_by_direct_integration}
  \widetilde{\delta_s} = 
  \int_0^\tau \widetilde{\omega}_{s\alpha} \mathrm{d}\tau',
  \label{eq:mdet2}
\end{equation}
with 
\begin{equation}
  \widetilde{\omega}_{s\alpha} =
  \frac{qI_p-I_t}{\Psi_p'} \left[ \tau_b^{-1} \partial_\alpha\rho_{\parallel s}
    - \overline{\tau_b^{-1} \partial_\alpha\rho_{\parallel s}} \right].
\end{equation}
Here, we have used the definition $\tau_b := B\sqrt{g}/(v_{\parallel}\Psi'_p)$ and the parameter $\tau$ given by
\begin{equation}
  \tau =
  \left\{
  \begin{array}{lcr}
    \int_{\theta_{b1}}^{\theta} |\tau_b|\, \mathrm{d}\theta'
    & \mathrm{when}\quad \sigma > 0 \vspace{0.2cm} \\
    \int_{\theta_{b1}}^{\theta_{b2}} |\tau_b|\, \mathrm{d}\theta 
    - \int_{\theta_{b2}}^{\theta} |\tau_b|\,
    \mathrm{d}\theta'  &  \mathrm{when} \quad \sigma < 0, 
  \end{array}
  \right.
\end{equation}
which is monotonic over the orbit. Finally, $\theta_{b1}$ and $\theta_{b2}$ are the bounce points of the orbit; that is, the solutions for $\theta$ of the equation $1-\lambda B(\psi,\theta,\alpha) = 0$. For the calculations of this paper we have implemented in CAS3D-K the expression (\ref{eq:delta_s_by_direct_integration}), instead of using an expansion in bounce harmonics as in \cite{Monreal2016}.

\subsection{Expression for the zonal flow oscillation frequency}
\label{sec:subsectionfrequencyexpression}

Using the initial condition\footnote{The only relevant feature of this initial condition, as far as the calculation of the zonal-flow frequency is concerned, is that $f_s(0) \approx (Z_s e / T_s) k_\perp^2\rho_{ts}^2 F_{s0}\varphi_k(0)$ for $k_\perp^2\rho_{ts}^2 \ll 1$.}
\begin{equation}
  f_s(0) = 
  \frac{Z_s e}{T_s} \frac{ \left\langle
    1-\Gamma_{0s}\right\rangle_\psi }{\Gamma_{0s}} 
  J_{0s} F_{s0}\, \varphi_k(0)
  \label{eq:ic}
\end{equation}
in (\ref{eq:gk6}), where $\Gamma_0\left(k_\perp^2\rho_{ts}^2\right) := \exp{\left(-k_\perp^2\rho_{ts}^2\right)} \,I_0\left(k_\perp^2\rho_{ts}^2\right)$ and $I_0$ is the zeroth order modified Bessel function, equation (\ref{eq:gk6}) reads
\begin{eqnarray}
  \widehat{\varphi}_k(p) = 
  \frac{
    \sum_s \frac{Z_s^2}{T_s} 
    \left\{ \frac{1}{p+\mathrm{i}k_\psi\overline{\omega_s}}
    \mathrm{e}^{-\mathrm{i}k_\psi\delta_s} J_{0s} 
    \overline{\mathrm{e}^{\mathrm{i}k_\psi\delta_s} 
      J_{0s} \left\langle 1-\Gamma_{0s}\right\rangle_\psi/ \Gamma_{0s} }
    \right\}_s
  }{
    \sum_s \frac{Z_s^2}{T_s} 
    \left\{1 - \frac{p}{p+\mathrm{i}k_\psi\overline{\omega_s}} \,
    \mathrm{e}^{-\mathrm{i}k_\psi\delta_s} J_{0s} 
    \overline{ \mathrm{e}^{\mathrm{i}k_\psi\delta_s} J_{0s}} \right\}_s  
  }
  \, \varphi_k(0).
  \label{eq:qn4}
\end{eqnarray}
The residual value of the electrostatic potential is obtained \cite{Monreal2016} from (\ref{eq:qn4}) by recalling the property of the Laplace transform $\lim_{t\to\infty}\varphi_k(t) = \lim_{p\to 0}p \widehat{\varphi}_k(p)$. Here, our goal is the derivation of an expression for $\Omega_{\rm ZF}$. This is found by expanding equation (\ref{eq:qn4}) in powers of $k_\psi \overline{\omega_s}/p \ll 1$, and keeping the lowest order terms in $k_\perp\rho_{ts} \sim k_\psi\delta_s \ll 1$. A direct check shows that, with these approximations,
\begin{equation}
  \widehat{\varphi}_k(p) = 
  \frac{ \varphi_k(0) }{  p\left(1+A_1/A_0\right) +p^{-1}A_2/A_0 },
  \label{eq:qn5}
\end{equation}
where
\begin{eqnarray}
  A_0 &=& \sum_s n_s \frac{Z_s^2}{T_s} \left\langle |\nabla \psi|^2 \rho_{ts}^2 \right\rangle_\psi, \label{eq:a0} \\
  A_1 &=& \sum_s \frac{Z_s^2}{T_s}  \left\{ \overline{\delta_s^2} - \overline{\delta_s}^2 \right\}_s  \label{eq:a1}
\end{eqnarray}
and 
\begin{eqnarray}
  A_2 &=& \sum_s \frac{Z_s^2}{T_s} \left\{ \overline{\omega_s}^2 \right\}_s.
  \label{eq:a2}
\end{eqnarray}
Taking the inverse Laplace transform of equation (\ref{eq:qn5}), we obtain the same result as in references \cite{Mishchenko2008,Helander2011}, namely
\begin{equation}
  \frac{\varphi_k(t)}{\varphi_k(0)} = \frac{1}{1 +A_1/A_0 } \cos\left( \Omega_\mathrm{ZF} t \right),
  \label{eq:qn6}
\end{equation}
where the zonal-flow frequency is given by
\begin{equation}
  \Omega_\mathrm{ZF} = \sqrt{\frac{A_2/A_0}{1+A_1/A_0}},
  \label{eq:zff}
\end{equation}
and the amplitude of the zonal-flow oscillation can be written as
\begin{equation}
  A_\mathrm{ZF} = \frac{1}{1+A_1/A_0}.
  \label{eq:zfa}
\end{equation}
Note that the expressions for the zonal-flow frequency and its amplitude do not depend on the radial wavenumber $k_\psi$, and that $A_\mathrm{ZF}$ is independent of the temperatures and the masses of the species.

From equation (\ref{eq:zff}), the estimate (\ref{eq:OmegaZFtypicalvalue}) is now understood. Clearly, a reduction of the bounce-averaged magnetic drift frequency, $\overline{\omega_s}$, leads to a decrease of the zonal-flow frequency, $\Omega_\mathrm{ZF}$. This allows to have larger values of the zonal flow in time scales comparable to those of the turbulence, as pointed out in \cite{Xanthopoulos2011}. Therefore, reducing $\overline{\omega_s}$ not only leads to a reduction of neoclassical transport but it might also be beneficial for the reduction of turbulent transport via zonal flows. This is one of the reasons that make the calculation of $\Omega_\mathrm{ZF}$ an interesting physical problem. The simultaneous optimization of neoclassical and turbulent transport has been addressed as well in \cite{Shaing2005,Sugama2005, Ferrando-Margalet2008,Watanabe2008}.

Whereas the frequency is a quantity that can be meaningfully compared with gyrokinetic simulations, the situation is different regarding the amplitude of the oscillation. The problem is that in practice the amplitude is damped \cite{Helander2011}, and in the analytical calculation above we have not considered the damping. For this reason, and due to the fact that $\Omega_{\rm ZF}$ seems to be more physically relevant, we focus on it in what follows.

We turn to the validity of the expression derived for $\Omega_{\rm ZF}$. The result (\ref{eq:zff}) is consistent with the orderings enumerated just before (\ref{eq:qn5}) because
\begin{equation}
  \frac{k_\psi \overline{\omega_s}}{\Omega_{\rm ZF}} \sim k_\perp \rho_{ts} \ll 1.
\end{equation}
If we recall (\ref{eq:conditions_validity_gkeq}), we deduce that (\ref{eq:zff}) is correct as long as
\begin{equation}\label{eq:conditions_validity_calculation_of_frequency}
  \frac{1}{L}\ll k_\perp \ll \frac{1}{\rho_{ts}},
\end{equation} 
as advanced in the Introduction.

The zonal-flow frequency (\ref{eq:zff}) and the amplitude (\ref{eq:zfa}) were first derived in \cite{Mishchenko2008} and the problem was analyzed in more detail in \cite{Helander2011}, where it was pointed out that (\ref{eq:conditions_validity_calculation_of_frequency}) (or, for what matters, (\ref{eq:validityapproximations})) might be difficult to satisfy in actual devices, and therefore the usefulness of (\ref{eq:zff}) was unclear in quantitative terms. In this work, we address a systematic comparison of the right side of (\ref{eq:zff}) with the value of $\Omega_{\rm ZF}$ obtained through gyrokinetic simulations. It turns out that (\ref{eq:zff}) is accurate (although there are a number of nuances about the assessment of the accuracy, depending on the device and associated to both the analytical calculation and the fitting of the frequency from gyrokinetic simulations, that are pointed out in the following sections) and that the numerical methods that we present to evaluate the right side of (\ref{eq:zff}) are faster than the determination of the frequency from gyrokinetic codes.

In the gyrokinetic simulations of this paper, plasmas consisting of one ion species and adiabatic electrons are employed. Then, in order to compare the semianalytical calculations with the results from gyrokinetic codes, the sum over species in equations (\ref{eq:a0}), (\ref{eq:a1}) and (\ref{eq:a2}) is then limited to one term, $s= i$. 

\section{Two kinds of numerical tools for the calculation of the zonal-flow oscillation frequency} 
\label{sec:Calculation}

The calculation of the zonal-flow frequency through (\ref{eq:zff}) involves the evaluation of the right sides of (\ref{eq:a0}), (\ref{eq:a1}) and (\ref{eq:a2}). In general, these expressions cannot be easily evaluated analytically. We use the code {\casdk} \cite{koenies2000,koenies2008,Monreal2016} to evaluate them numerically. In order to prove the accuracy of (\ref{eq:zff}) to determine the oscillation frequency in actual stellarator configurations, we compare its evaluation with {\casdk} against the frequency obtained from simulations with the radially global gyrokinetic code {\eut} \cite{Jost2001,Kleiber2012} and the radially local, full flux-surface version of {\gene} \cite{Jenko2000,Gorler2011,GENE,Xanthopoulos2009}.

The code {\casdk} allows to perform averages of phase-space functions over the lowest order trajectories. As these trajectories lie on flux surfaces, the averages involve basically two-dimensional integrations over the spatial coordinates. We use $\{\psi,\theta,\zeta\}$ as independent spatial coordinates for passing particles and $\{\psi,\theta,\alpha\}$ for trapped particles, and $\{v,\lambda,\sigma\}$ as independent velocity space coordinates. In particular, the averaging operations implemented in {\casdk} are the orbit-average, described in equation (\ref{eq:bounceaverages}), and the phase-space average operation which is given in (\ref{eq:corchete}). A more detailed description of how these averages are performed in {\casdk} can be found in reference \cite{Monreal2016}. In this work, the calculation of $\delta_s$ for trapped particles has been significantly simplified in {\casdk} by using the direct integration given by equation (\ref{eq:mdet2}) instead of the Fourier method originally implemented in {\casdk} and described in \cite{Monreal2016}. The values of  $\delta_s$ obtained with both methods are the same while the new one is much simpler to implement and faster.

{\eut} is a global $\delta f$ gyrokinetic code in 3D geometry with a Lagrangian Particle In Cell (PIC) scheme. The gyrokinetic simulations carried out with {\eut} in this work are linear and collisionless with a plasma that consists of singly charged ions and adiabatic electrons with equal temperatures. The simulations are initiated with a zonal potential perturbation, which is produced by taking a perturbed distribution function for the ions
\begin{equation}
  F_{i1}(0) = \epsilon 
  \frac{e}{T_i} \left\langle k_\perp^2\rho_{ti}^2 
  \right\rangle_\psi \varphi_k(0) \cos(k_\psi \psi)\, F_{i0}.
  \label{eq:initialeuterpe}
\end{equation}
Here, $F_{i0}$ is a Maxwellian distribution and $\epsilon$ is a small factor (on the order of $10^{-3}$) that makes the perturbation to the distribution function, $F_{i1}$, much smaller than the equilibrium distribution, $F_{i0}$.

A long-wavelength approximation is used in the quasineutrality equation in this work. Namely, the  $\Gamma_0(x)=\mathrm{e}^{-x}I_0(x)$ function is approximated as $\Gamma_0(x)\approx 1- x$, with $x=k_{\perp}^2\rho_{ti}^2$ and $I_0$ being the modified Bessel function, which is valid for  $k_\perp\rho_{ti} \ll 1$. The initial condition (\ref{eq:initialeuterpe}) is a good approximation of (\ref{eq:ic}) under these conditions. Note also that the simulation is carried out with a fixed value of $k_{\psi}$, so that $k_{\perp}\rho_{ti}$ has a radial dependence due to the radial variation of the magnetic field. The radial variation of $k_{\perp}\rho_{ti}$ is not relevant, however, because the oscillation frequency does not depend on the radial scale (see Section~\ref{sec:Evolution}). This was confirmed by carrying out several simulations for different values of $k_{\psi}$, from $0.5\pi$ to $4.5\pi$, with the same configuration and same numerical parameters.

With the initial condition (\ref{eq:initialeuterpe}) for ions, the collisionless simulation is linearly evolved, retaining just a few ($5$ to $10$) toroidal and poloidal Fourier modes of the potential. The $m=0, n=0$ mode, which we can identify with $\varphi_{k}$ in equation (\ref{eq:eikonalvarphi}), is the dominant component of the potential spectrum during the simulation, thus proving that the assumption of an eikonal form in equation (\ref{eq:eikonalvarphi}) is appropriate. Here, $m$ and $n$ label poloidal and toroidal modes, respectively. The time evolution of $\varphi_{k}$  at a number of radial positions is tracked.

To obtain the frequency of the zonal-flow oscillation, the time trace of the zonal potential, or its radial derivative\footnote{Assuming the eikonal form of the potential in equation (\ref{eq:eikonalvarphi}), both the normalized potential and its radial derivative have a similar time evolution, with the same oscillation frequency.} (normalized to its initial value), is fitted to a model function with the form
\begin{equation}
  \varphi'_{k}(t)/\varphi'_{k}(0) = 
  A_\mathrm{ZF} \cos(\Omega_\mathrm{ZF} t) \exp{(-\gamma_\mathrm{ZF} t)} 
  + R_\mathrm{ZF} 
  + \frac{c}{1+ d t^e},
  \label{eq:fit}
\end{equation}
where $\Omega_\mathrm{ZF}$ is the zonal-flow oscillation frequency, $A_\mathrm{ZF}$ is the amplitude of the oscillation, $\gamma_\mathrm{ZF}$ is the damping rate and $R_\mathrm{ZF}$ is the residual level, and $'$ means derivative with respect to the radial coordinate (normalized toroidal flux in {\eut}). The last term in equation (\ref{eq:fit}) accounts for the decay to the zonal-flow residual level, to which the oscillations are superimposed. This term is important for cases in which the decay to this residual level is slow; e.g. in tokamak configurations with a small ripple added (see Section~\ref{sec:Tokamakresults}). The fit is performed with the non-linear fitting routine \textit{fit} of the {\small MATLAB} software package. Note that, in principle, a similar value of the zonal-flow frequency could be obtained from a FFT of the time signal, but this method shows to be less precise in practice.

There are sources of error and uncertainty in the oscillation frequency obtained by this procedure, related to the simulation itself and the approximations in the gyrokinetic code, and also associated to the fitting process. Several dynamics are mixed in the zonal flow relaxation, which makes the fitting process non trivial. First, there is a decay to the residual level and, superimposed to it, several oscillations appear with different amplitudes and characteristic times (see figures~\ref{fig:t-tok} and \ref{fig:t-stell}). Since we are interested in the low frequency oscillation, the faster GAM oscillation, usually appearing at the beginning of the relaxation, is excluded in the fitting process. A very similar value is obtained if instead of using the $m=0,n=0$ component of the potential its flux-surface average is used\footnote{Note that the $m=0,n=0$ component of the potential is not strictly equal to the flux-surface average of the potential, in general.}.

Although the comparison of the frequency calculations is focused on the codes {\casdk} and {\eut}, calculations for the W7-X standard configuration have also been carried out with the code {\gene}. This is an Eulerian gyrokinetic $\delta f$ code which can be run in radially global, full flux surface or flux tube simulation domains. In the {\gene} simulations shown in this work, we use the full-flux surface version and adiabatic electrons. In this version the {\gene} code is spectral in the radial coordinate, while a finite difference scheme is used for the coordinates along the flux surface. The distribution function is initialized  according to (\ref{eq:ic}). From the {\gene} simulations, as in the case of {\eut}, the frequency is obtained by fitting the normalized zonal potential time trace to a model like (\ref{eq:fit}).

In all cases we start from a magnetic equilibrium calculated with the 3D MHD equilibrium code {\small VMEC} \cite{Hirshman1983}. Then, the relevant equilibrium quantities are mapped from {\small VMEC} to PEST coordinates used in {\eut}  by means of an intermediate code; from {\small VMEC} to {\gene} coordinates by means of the GIST \cite{Xanthopoulos2009} package; and from {\small VMEC} to Boozer coordinates used in {\casdk} by employing the MC3D code which is part of the {\small CAS3D} \cite{Schwab1993} code package, and also part of {\casdk}.

For convenience, we use flat density and temperature profiles and assume a plasma with adiabatic electrons and singly charged ions in all cases. Although the normalized toroidal flux, $\psi$, is used as a radial coordinate in the codes, in what follows we show the results in the more natural coordinate $r/a := \sqrt{\psi}$, where $a$ is the minor radius of the device.

\section{Zonal-flow oscillation frequency in tokamaks with ripple}
\label{sec:Tokamakresults}

In order to understand the influence of the magnetic configuration on the zonal-flow oscillation, we take an axisymmetric equilibrium and modify it by adding different amounts of ripple. The resulting set of configurations allows us to study the dependence of the frequency on the ripple size and calibrate the calculation methods in a controlled path continuously departing from the axisymmetric case. As we explain below, the axisymmetric tokamak configuration employed here is completely academic because we  use a safety profile that is not realistic. The reason for this is that, in this section, we are interested in showing in a pedagogical way how $\Omega_{\rm ZF}$ increases as the magnitude of the axisymmetry-breaking terms grows, both using the semianalytical method and gyrokinetic simulations. Regarding the latter, the safety profile chosen helps remove the GAM oscillation, and therefore allows to clearly observe the zonal flow oscillation and to measure its frequency.

\begin{figure}
    \includegraphics[width=1.\linewidth]{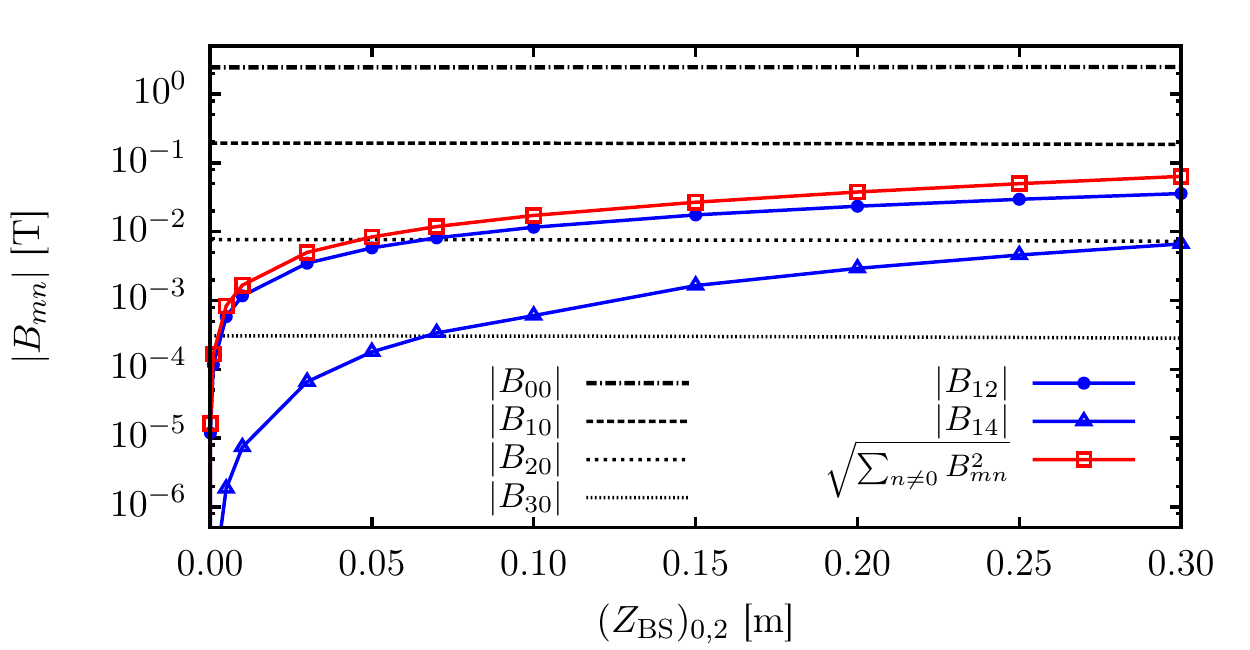}
  \caption{Some relevant quantities that describe the rippled tokamak configurations studied in the text as a function of the non-axisymmetric perturbation to the plasma boundary, $(Z_{\rm BS})_{0, 2}$. The values in the figure correspond to those quantities evaluated at $r/a=0.8$.}
  \label{fig:ModosBTokSeries}
\end{figure}

We start with the {\small VMEC} input for the equilibrium of an axisymmetric large aspect ratio tokamak (LART) with major radius $R=5$~m and minor radius $a=0.5$~m, and modify the plasma boundary to include ripple. The ripple is generated by adding a poloidally symmetric perturbation to the boundary shape through non-zero coefficients $(R_{\rm BC})_{0,2}$ and $(Z_{\rm BS})_{0, 2}$. We always take $(R_{\rm BC})_{0, 2} = (Z_{\rm BS})_{0, 2}$. Here,  $(R_{\rm BC})_{m,n}$ and $(Z_{\rm BS})_{m,n}$ are the cosine and sine components of the coordinates $R_{\rm BC}$ and $Z_{\rm BS}$ at the boundary, respectively, and $m$ and $n$ are poloidal and toroidal mode numbers (see \cite{Lazersonwww}). We calculate the {\small VMEC} equilibrium employing 36 Fourier cosine components\footnote{$|B|$ can be written as a Fourier cosine series with coefficients $B_{m,n}$ as $|B|=\sum_{m=0}^{3}\sum_{n=-4}^4 B_{m,n}\cos(m\theta-n  \zeta)$.} for $B$; specifically, $0\leq |n| \leq 4$  and $0 \leq m \le 3$. After including the $(Z_{\rm BS})_{0, 2} \ne 0$ perturbation to the boundary, several modes $B_{0,n}$, $n\ne 0$, are generated, whose size increases with the size of $(Z_{\rm BS})_{0,2}$. The non-axisymmetric perturbation also modifies, although only slightly, the axisymmetric components $B_{00}$, $B_{10}$, $B_{20}$ and $B_{30}$.  We show in figure \ref{fig:ModosBTokSeries} some of the Fourier cosine coefficients of $B$ at the radial position $r/a=0.8$ for different values of the perturbation $(Z_{\rm BS})_{0, 2}$. In the figure, we give  the axisymmetric components, the largest non-axisymmetric coefficients, $B_{12}$ and $B_{14}$, and also $\sqrt{\sum_{n\ne0} B_{mn}^2}$, where the sum includes all non-axisymmetric components. For the cases under study, we define the ripple by
\begin{equation}
  \xi = \frac{\sqrt{\sum_{n\ne0} B_{mn}^2}} { B_{10}}.
\end{equation}
This parameter will be used as a measure of the perturbation to the axisymmetric equilibrium.

\begin{figure}
  \includegraphics[width=1.\linewidth]{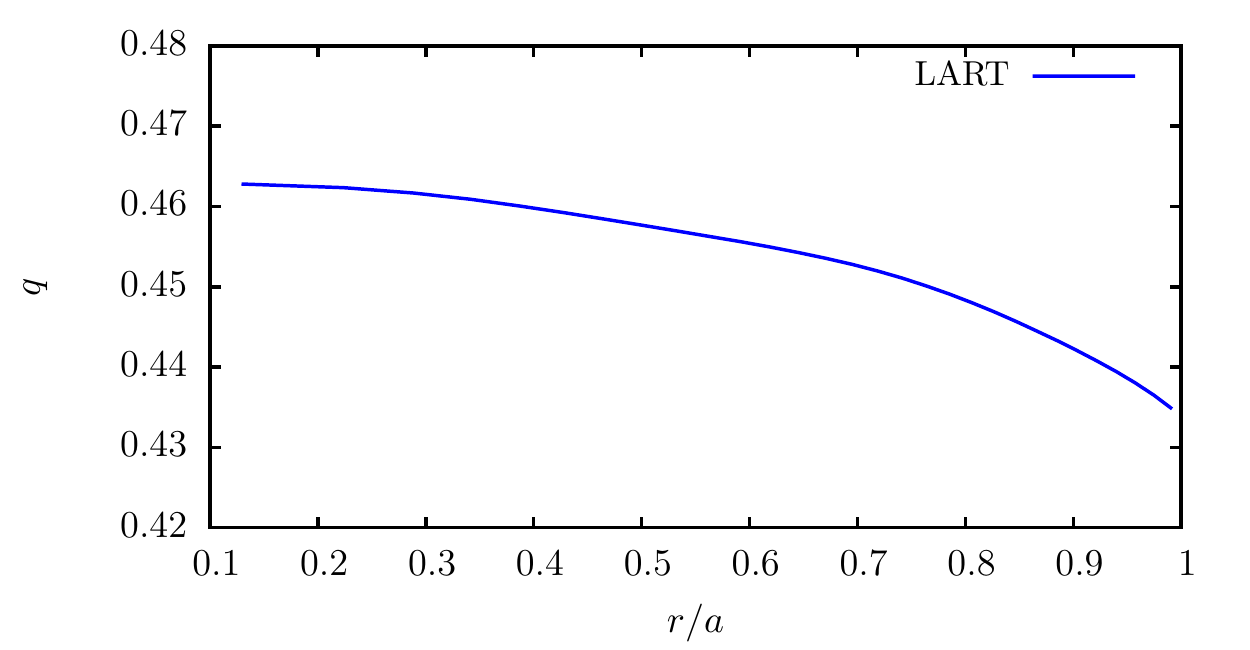}
  \caption{Radial profile of the safety factor of the large aspect ratio tokamak (LART) equilibria described in Section~\ref{sec:Tokamakresults}.}
  \label{fig:q-tok}
\end{figure}

The average magnetic field strength at the  axis is $B_0\simeq 2.48$~T for the non-perturbed case and decreases slightly as the ripple amplitude is increased in the perturbed equilibria. The $q$ profile, which is the same for all these configurations, is shown in figure~\ref{fig:q-tok}. This $q$ profile is not standard for a tokamak configuration. Actually, it corresponds to a TJ-II configuration with high rotational transform. The collisionless Landau damping of the GAM oscillation strongly depends on the safety factor \cite{Sugama2006b}. Therefore, as advanced above, we use unusually small safety factor values to reduce the amplitude of the GAM oscillation, which allows to observe the low frequency oscillation. For a $q$ profile typical from tokamaks, the GAM is weakly damped even at the center where $q$ is small, and the low frequency oscillation in which we are interested is undetectable even for large values of ripple added.

In this set of configurations we calculate the zonal flow oscillation frequency with {\casdk} and also by means of EUTERPE simulations. We use flat density and temperature profiles with $T_e=T_i=5$~keV. In {\eut} we use an initial perturbation with $k_{\psi}=0.5 \pi$, so that the normalized radial scale of the perturbation is $\left\langle k_{\perp} \rho_{ti}\right\rangle_\psi < 0.026$.

\begin{figure}
  \includegraphics[width=1.\linewidth]{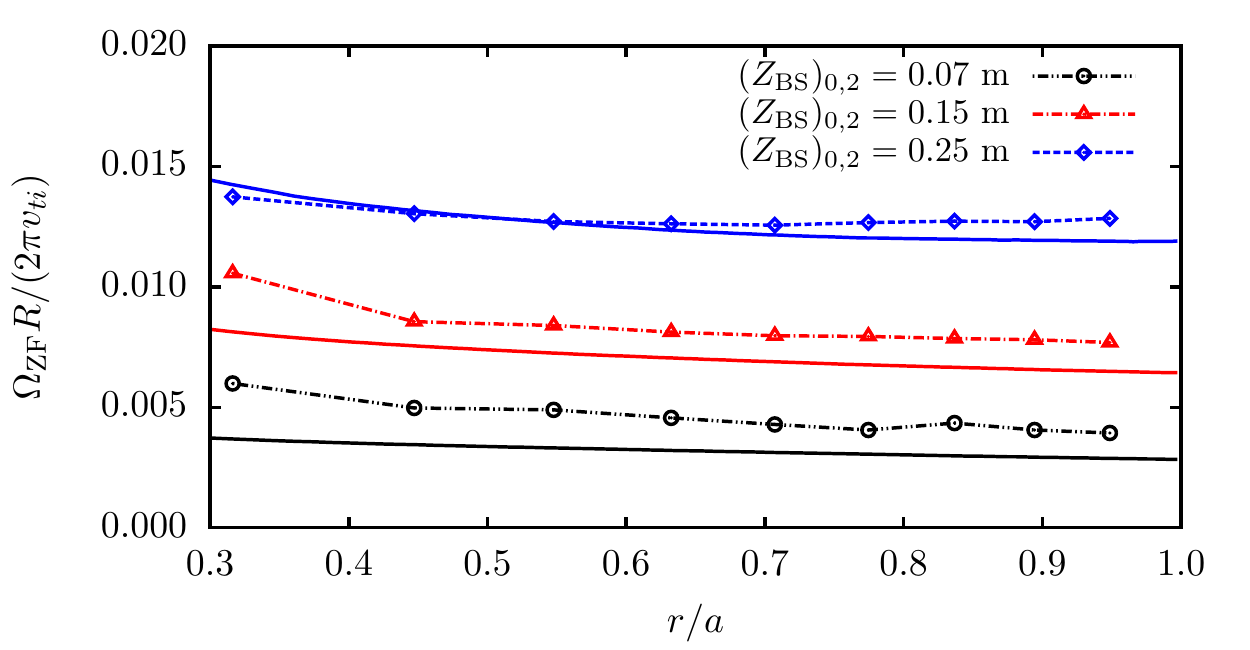}
  \caption{Radial dependence of $\Omega_{\rm ZF}$ in the large aspect ratio tokamak with different ripple values. The results of EUTERPE are shown with dashed lines (the specific points of the radial grid are marked) and the results of CAS3D-K are shown with solid lines.}
  \label{fig:r-tok}
\end{figure}

\begin{figure}
  \includegraphics[width=1.\linewidth]{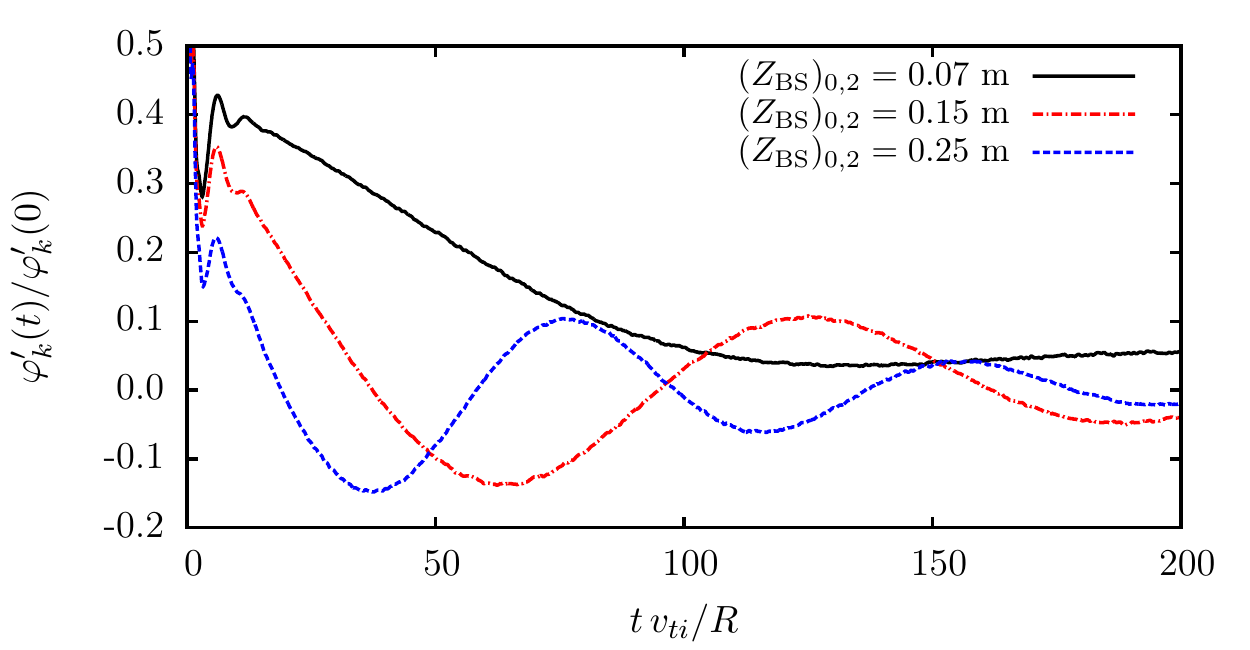}
  \caption{Time evolution of the normalized radial electric field in the large aspect ratio tokamak with different ripple values obtained with EUTERPE at $r/a=0.5$.}
  \label{fig:t-tok}
\end{figure}

The radial dependence of the zonal-flow frequency calculated with {\casdk} and {\eut} for the tokamaks described above is shown in figure~\ref{fig:r-tok} for different ripple values. The oscillation frequency shows rather flat radial profiles with a slight increase near the magnetic axis. The calculations with both codes show good agreement and this is better in configurations with larger ripple amplitudes; the reason is given next. The time traces of the normalized zonal electric field corresponding to the cases shown in figure~\ref{fig:r-tok} are provided in figure~\ref{fig:t-tok} at a radial position of $r/a=0.5$. As can be seen in this figure, the initial time steps are dominated by a fast oscillation followed by a decaying smaller-frequency oscillation which has larger amplitude in the tokamak configurations with larger ripple values. It is then clear from figure~\ref{fig:t-tok} that the estimation of the frequency with the fitting method is less precise in configurations with smaller ripple values because the amplitude of the zonal-flow oscillation in those configurations is actually small and it is superimposed to a slow decay to the residual level. In particular, the fitting method fails for sufficiently small $\xi$. Hence, the slight differences in the calculation of $\Omega_{\rm ZF}$ from the semianalytical approach and from the gyrokinetic simulations seem to be simply due to the uncertainties in the fitting method.

\begin{figure}
  \includegraphics[width=1.\linewidth]{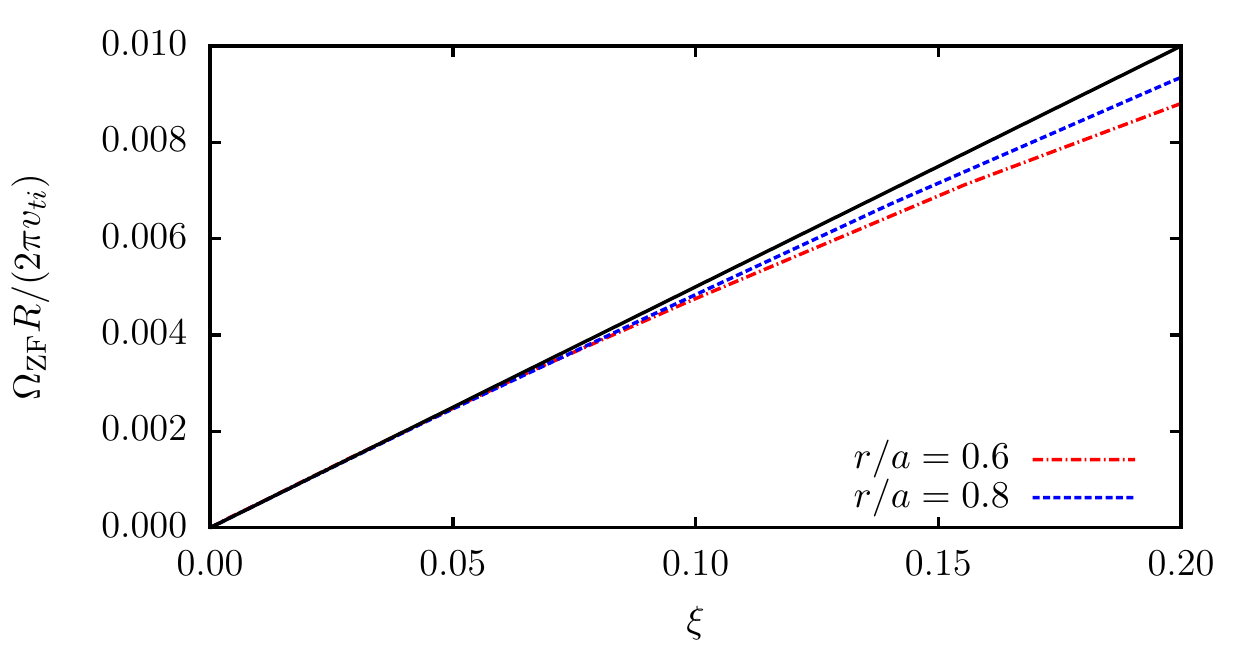}
  \caption{Dependence of $\Omega_{\rm ZF}$ with $\xi$ in the large aspect ratio tokamak described in Section~\ref{sec:Tokamakresults}. Observe that for $\xi \ll 1$ the dependence is linear. The black straight line is a linear fit of the other two curves around $\xi = 0$.}
  \label{fig:b_tok_c3k}
\end{figure}

With CAS3D-K we can compute $\Omega_{\rm ZF}$ for arbitrarily small values of the ripple. Theoretically, a linear dependence of $\Omega_{\rm ZF}$ on $\xi$ is expected for $\xi \ll 1$. Note that the results of \cite{Calvo13} are applicable here and imply $\overline{\omega_i}\sim \xi \rho_{ti}v_{ti}/L$ for $\xi \ll 1$. Since $A_0$ and $A_1$ (see (\ref{eq:a0}) and (\ref{eq:a1})) are dominated by the axisymmetric terms, the expansion of (\ref{eq:zff}) for $\xi \ll 1$ yields
\begin{equation}
  \Omega_{\rm ZF} \sim \xi \frac{v_{ti}}{L}.
\end{equation}
This is confirmed by figure~\ref{fig:b_tok_c3k}.

The series of perturbed tokamak equilibria used in this section have allowed us to clarify the dependence of the oscillation frequency with the ripple and understand some difficulties that appear in the process of comparing the two approaches that are the subject of this paper. In particular, the calculation of the oscillation frequency by means of gyrokinetic simulations showed to have problems when the ripple size is too small because the fit of the potential time traces becomes complicated. The calculation with {\casdk}, however, does not have this limitation and proved to be robust even with very small ripple amplitudes.

\section{Zonal-flow oscillation frequency in stellarators}
\label{sec:Stellaratorresults}

In this section, we calculate the zonal-flow frequency in stellarator geometries with {\casdk} and compare the results with those obtained from gyrokinetic simulations using the codes {\eut} and {\gene}. Specifically, we work out the frequency in several magnetic configurations of the W7-X, TJ-II and LHD stellarators.

\begin{figure}
  \includegraphics[width=1.\linewidth]{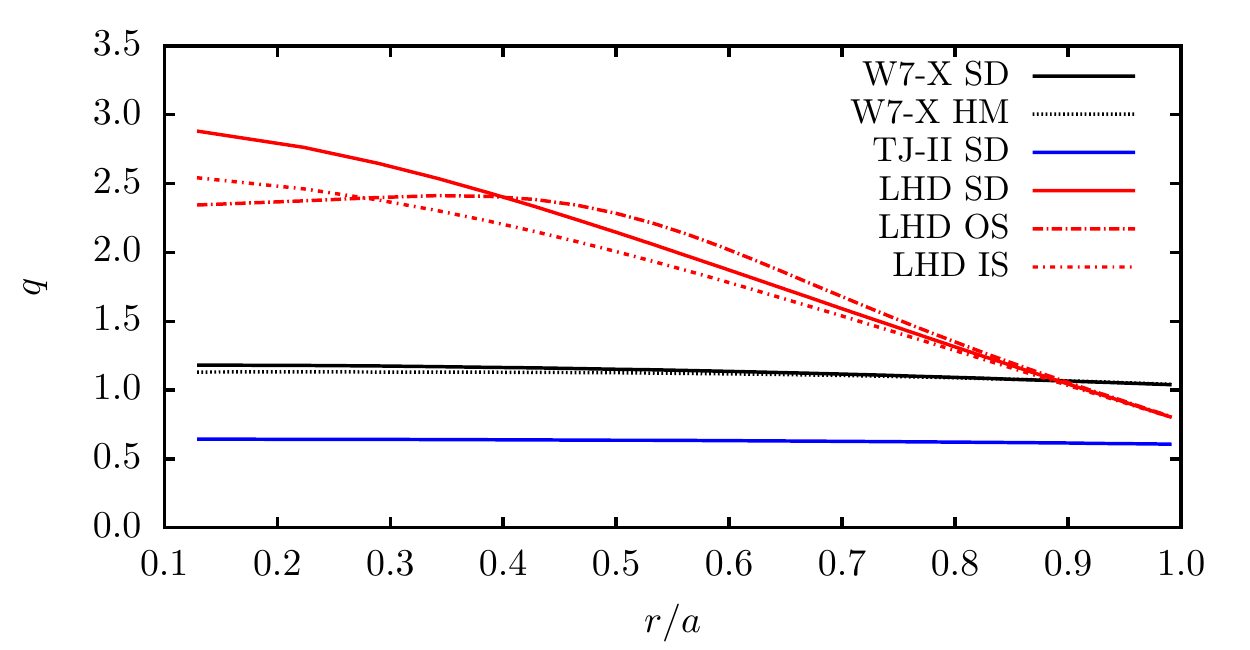}
  \caption{Safety factor radial profiles of the stellarator configurations described in Section~\ref{sec:Stellaratorresults}.}
  \label{fig:q-all}
\end{figure}

\begin{figure}
  \includegraphics[width=1.\linewidth]{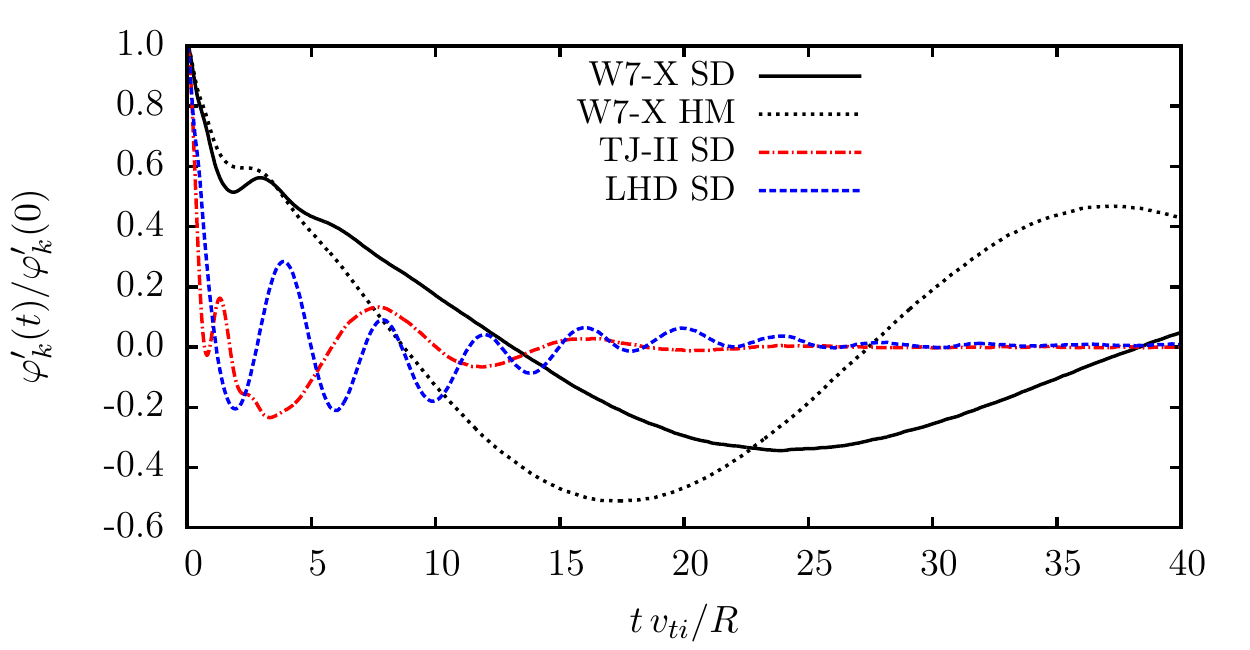}
  \caption{Time evolution of the normalized radial electric field obtained with EUTERPE for some of the stellarator configurations studied in Section~\ref{sec:Stellaratorresults} at $r/a=0.5$. For the normalization of time in the horizontal axis, we have employed the major radius of each stellarator and the values of the thermal speed given in the corresponding subsection of Section~\ref{sec:Stellaratorresults}. It is important to emphasize that, for the LHD SD curve, the oscillation that is seen with the naked eye does not correspond to the low-frequency oscillation but to the GAM.}
  \label{fig:t-stell}
\end{figure}

As discussed in previous sections, the time evolution of an initial zonal-flow perturbation shows, in general, an initial damped GAM oscillation followed by a decaying low frequency oscillation. The safety factor profiles for all the stellarator configurations studied in this section are shown in figure~\ref{fig:q-all}. An example of the time traces of the zonal electric field obtained with {\eut} is given in figure~\ref{fig:t-stell} for some of the considered stellarator configurations. As can be seen in this figure, the evolution of the initial zonal-flow perturbation is qualitatively different in each device. The standard and high mirror configurations of W7-X show a very small amplitude GAM oscillation before $t=5 R/v_{ti}$ and a much slower and higher-amplitude oscillation afterwards. The curve corresponding to LHD shows up to six clear cycles of the GAM oscillation, while the low frequency one is almost imperceptible. TJ-II represents an intermediate situation:  the amplitude of the GAM oscillation is larger than in W7-X but smaller than in LHD. As for the low frequency oscillation, two oscillation cycles are perfectly observable. Therefore, obtaining an accurate value of the zonal-flow frequency from the fitting method requires a careful analysis case by case.

\subsection{W7-X stellarator: standard configuration}
\label{sec:w7xresults}

\begin{figure}
  \includegraphics[width=1.0\linewidth]{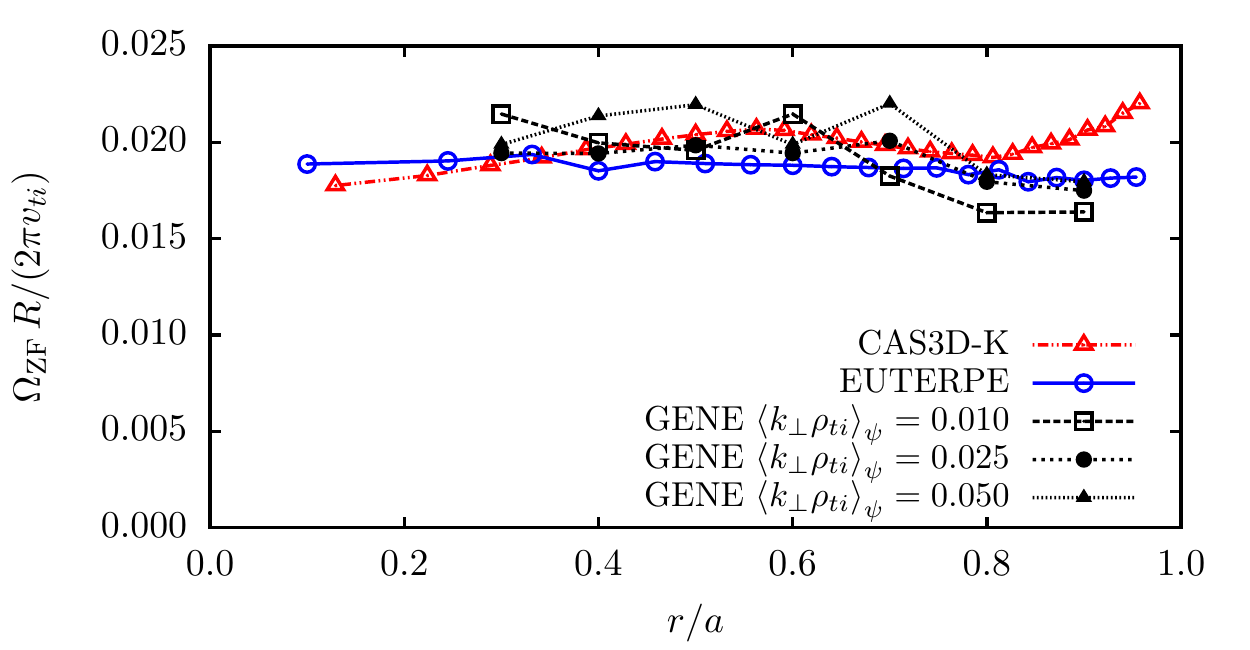}
  \caption{Zonal-flow frequency in the standard configuration of the W7-X stellarator obtained with CAS3D-K, EUTERPE and GENE. Several values of $\left\langle k_\perp\rho_{ti}\right\rangle_\psi$ are shown for GENE calculations. The EUTERPE calculations use $k_{\psi}=0.5\pi$.}
  \label{fig:zff-w7x-ceg}
\end{figure}

In this subsection we calculate the oscillation frequency in the standard configuration of the W7-X stellarator (W7-X SD). In the W7-X SD configuration the average magnetic field strength at the magnetic axis is $B_0=2.42$~T. The $q$ profile of this configuration is given in figure~\ref{fig:q-all}. In the calculations presented here we use flat density and temperature profiles with $T_e=T_i=5$~keV. 

The calculations of the zonal-flow frequency with {\casdk}, {\eut} and {\gene} in the W7-X SD configuration are shown in figure~\ref{fig:zff-w7x-ceg}. The oscillation frequency in all cases shows a rather flat radial profile and the agreement between them is very good. In the simulations carried out with {\eut} we use an initial perturbation with $k_{\psi}=0.5\pi$, so that the normalized radial scale of the perturbation varies radially, with $\left\langle k_{\perp} \rho_{ti}\right\rangle_\psi < 0.032$. Very similar values of the oscillation frequency were obtained in simulations with different (small) values of $k_{\psi}$ (not shown here). Analogously, several simulations were carried out with {\gene} for different radial scales of the perturbation with $0.01\leq \left\langle k_\perp\rho_{ti}\right\rangle_\psi\leq0.05$ obtaining very similar values of the oscillation frequency (see figure \ref{fig:zff-w7x-ceg}). These results are in agreement with equation (\ref{eq:zff}), which is independent of $k_{\perp}$ in the long-wavelength limit. 

In the {\eut} calculations, the error when obtaining the frequency is larger in the outer region of the plasma because the fit to the model function gives a less precise value. The treatment of lost particles at the outer boundary can also introduce some bias in the oscillation frequency. In the {\eut} calculations an error on the order of a $20\%$ of its value can be assumed in these radial positions. It is in these radial positions where the {\casdk} calculation shows a significant increase in the frequency and the differences with the {\eut} calculations are on the order of the error.

Finally, it might be useful to provide values for the frequency in physical units.  Going back to figure \ref{fig:zff-w7x-ceg}, recalling that in this case $T_i=5$~keV and employing that the major radius of W7-X is $R=5.5$~m, we find $\Omega_{{\rm ZF}}/(2\pi) \sim 2.5$~kHz.

\subsection{TJ-II stellarator}
\label{sec:tj2results}

We have calculated the zonal flow oscillation frequency in the standard configuration of the TJ-II stellarator with {\casdk} and {\eut}. The $q$ profile of this configuration is given in figure \ref{fig:q-all}. We also use flat density and temperature profiles, in this case with temperatures $T_e=T_i=100$~eV, close to typical ion temperature values in ECRH plasmas of TJ-II. In {\eut} we use, as in the previous case, an initial perturbation with $k_{\psi}=0.5\pi$, so that the normalized radial scale of the perturbation is $\left\langle k_{\perp} \rho_{ti}\right\rangle_\psi < 0.023$. 

\begin{figure}
  \includegraphics[width=1.\linewidth]{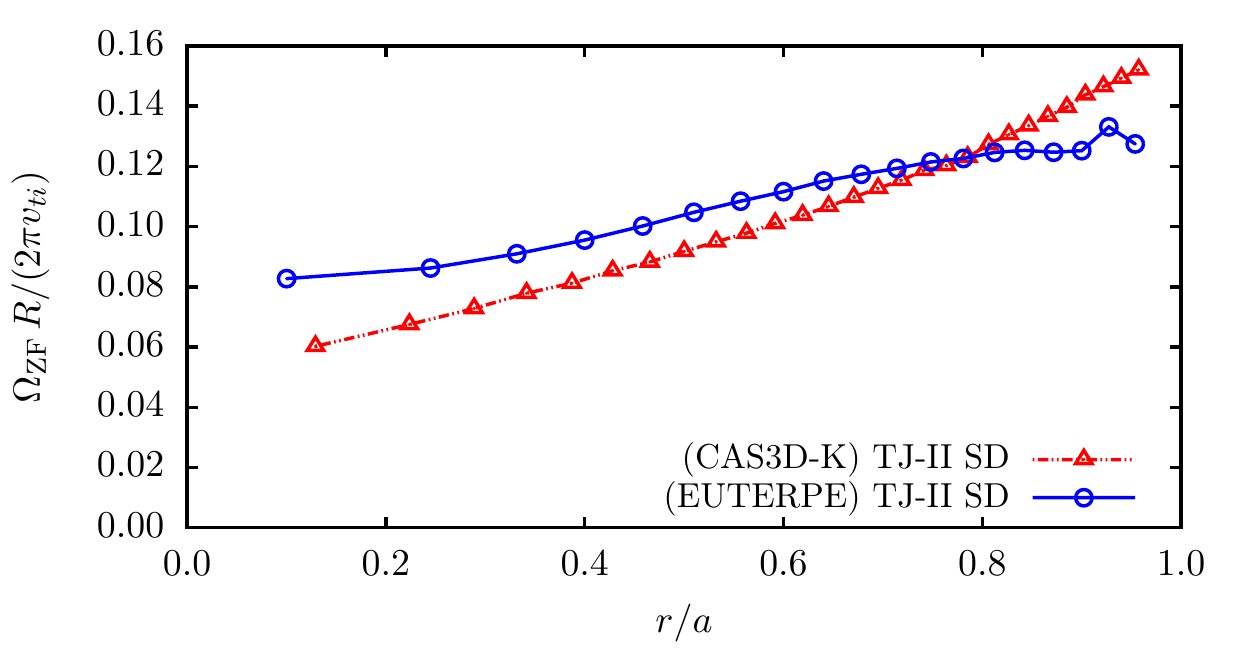}
  \caption{Calculation of the zonal-flow frequency in the standard configuration of the TJ-II stellarator obtained with CAS3D-K and EUTERPE. The radial scale of the initial perturbation in EUTERPE simulations is $k_{\psi}=0.5\pi$.}
  \label{fig:zff-tj2}
\end{figure}

The results of the zonal-flow frequency calculated with {\casdk} and {\eut} for this configuration are shown in figure \ref{fig:zff-tj2}. As can be seen in this figure, the zonal-flow frequency in the TJ-II stellarator increases with the radial coordinate $r/a$ and the calculations with the different numerical methods show a good agreement. 

It is important to point out that in this configuration the calculation of $\Omega_{\rm ZF}$ from the fit to the model function (\ref{eq:fit}) is not as accurate as in the W7-X SD case because the damping of the low frequency oscillation is larger in TJ-II and fewer oscillation cycles are observed (see figure~\ref{fig:t-stell}). However it can be calculated with reasonable accuracy from the {\eut} electric field time traces  for all radial positions because the initial GAM oscillation is quickly damped leaving a couple of cycles of neat zonal-flow low frequency oscillation. The first part of the time trace, in which the GAM oscillation is present, is not included in the fit.

From figure \ref{fig:zff-tj2} we can deduce that in TJ-II, whose major radius is $R=1.5$~m, the frequency varies radially in the range $\Omega_{{\rm ZF}}/(2\pi) \sim 5-8$~kHz for $T_i=100$~eV.

\subsection{LHD stellarator}
\label{sec:lhdresults}

\begin{figure}
  \includegraphics[width=1.0\linewidth]{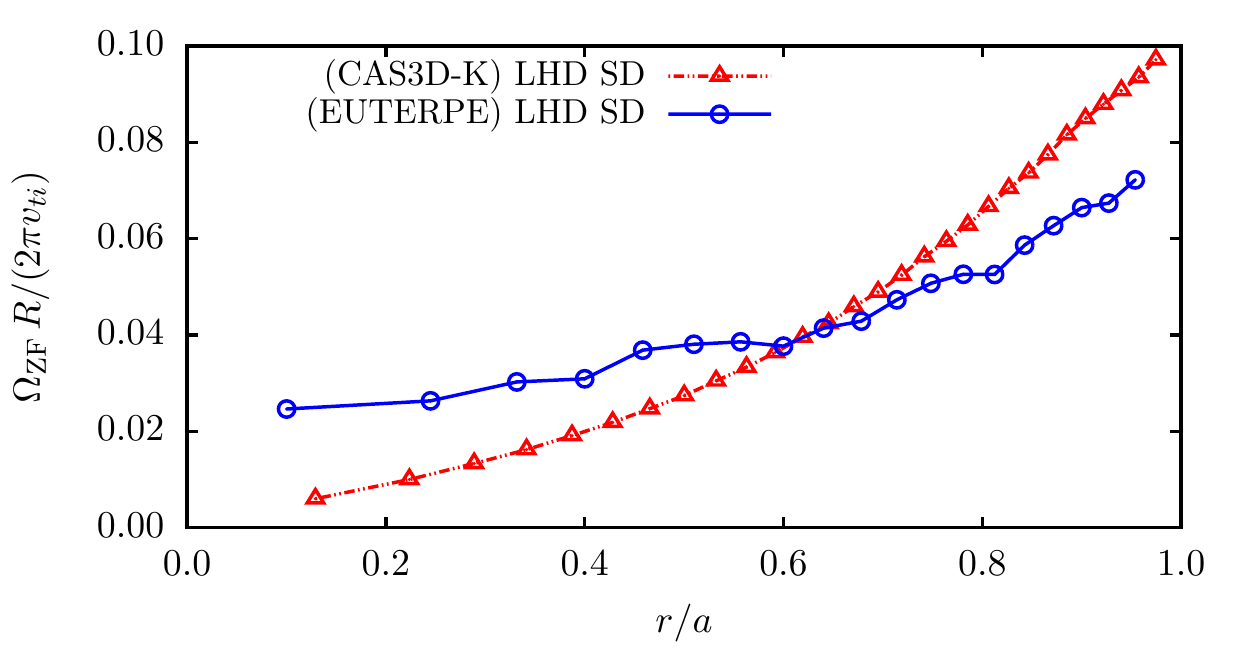}
  \caption{Zonal-flow frequency in the standard configuration of the LHD stellarator obtained with CAS3D-K and EUTERPE.}
  \label{fig:zff-lhd0}
\end{figure}

In the LHD stellarator, we calculate the zonal-flow frequency in three magnetic configurations with different positions of the magnetic axis, $R_\mathrm{ax}$, and different values of the magnetic field strength at the magnetic axis, $B_0$. These are: the standard configuration (LHD SD) with $R_\mathrm{ax}=3.74$~m and $B_0=2.53$~T; an outward-shifted configuration (LHD OS) with $R_\mathrm{ax}=3.91$~m and $B_0=1.48$~T; and an inward-shifted configuration (LHD IS) with $R_\mathrm{ax}=3.57$~m and $B_0=1.57$~T. The $q$ profiles of these configurations are given in figure~\ref{fig:q-all} and we take flat density and temperature profiles with $T_i = T_e=5$~keV. In {\eut} simulations we use an initial perturbation with $k_{\psi}=0.5\pi$, so that the normalized radial scale of the perturbation is $\left\langle k_{\perp} \rho_{ti}\right\rangle_\psi < 0.018$.

The frequencies calculated with {\casdk} and {\eut} in the LHD SD configuration are shown in figure~\ref{fig:zff-lhd0}. Both  calculations show a clear radial increase of the frequency. In this case, the agreement between calculations with {\casdk} and {\eut} cannot be expected to be optimal, as can be easily understood from a simple inspection of  figure~\ref{fig:t-stell}. This figure shows that the GAM oscillation is weakly damped and it is almost impossible to distinguish the low frequency oscillation. Therefore, in this configuration the fitting process turns out to be very complicated. In this sense, the agreement shown in figure~\ref{fig:zff-lhd0} is actually remarkable.

\begin{figure}
  \includegraphics[width=1.0\linewidth]{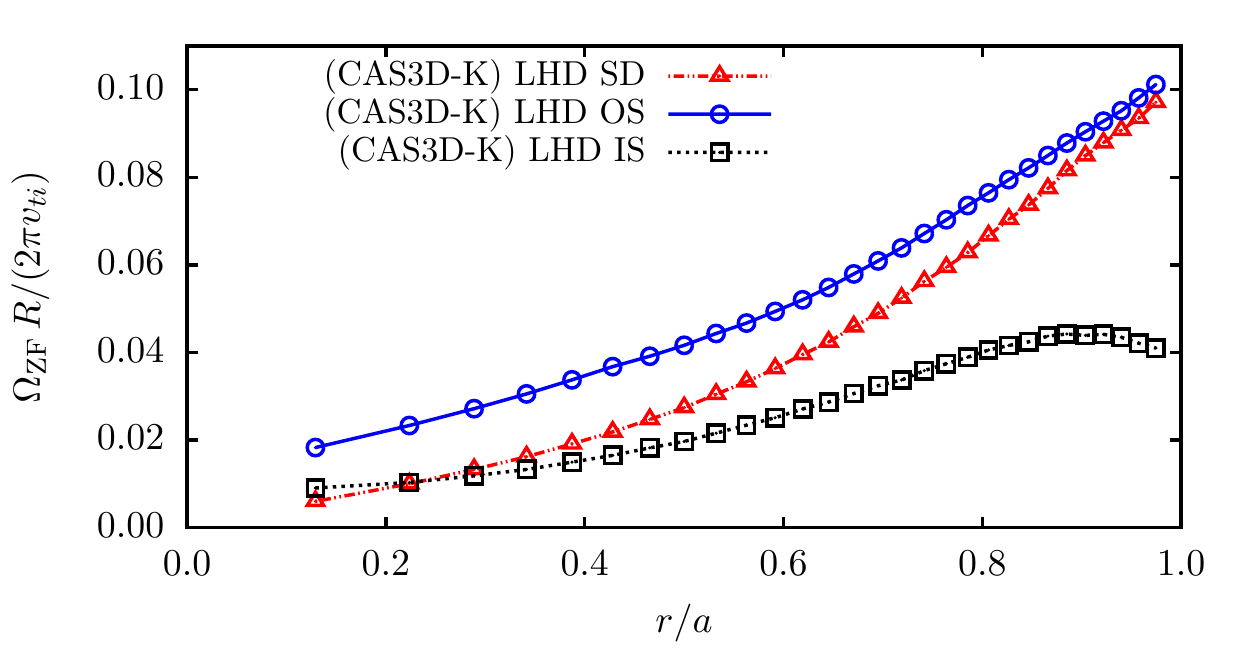}
  \caption{Zonal-flow frequency obtained with CAS3D-K in the LHD standard (LHD SD), outward-shifted (LHD OS) and inward-shifted (LHD IS) configurations.}
  \label{fig:zff-lhd-c3k}
\end{figure}

In figure \ref{fig:zff-lhd-c3k} we show the {\casdk} calculations of the zonal-flow frequency in the LHD SD, LHD OS and LHD IS configurations. They also exhibit a radial increase of the frequency in all configurations, except for the LHD IS configuration in external radial positions, for $r/a>0.9$. The inward-shifted configuration is better optimized for neoclassical transport, which has lower values of the averaged magnetic drift frequency, $\overline{\omega_s}$. This, at the same time, shows a larger value of the zonal-flow residual level (see reference \cite{Watanabe2008}). As can be seen in figure~\ref{fig:zff-lhd-c3k}, the frequency of the zonal-flow oscillation in the LHD IS configuration is smaller than in the LHD SD or LHD OS configurations, which is considered also a direct consequence of the reduction in $\overline{\omega_s}$. 

As we did for the stellarator configurations discussed in previous subsections, we give values of the oscillation frequency in physical units for the LHD configurations studied here. Recall that $T_i = 5$~keV. In the standard and outward shifted configurations the frequency increases radially in the range $\Omega_{\rm ZF}/(2\pi) \sim 1.85- 18.5$~kHz, while in the inward shifted configuration it lies in the range $\Omega_{\rm ZF}/(2\pi) \sim 1.9- 8.5$~kHz according to CAS3D-K calculations.

We do not show the calculation of $\Omega_{\rm ZF}$ with {\eut} in the IS and OS configurations of LHD because a fit to a model like (\ref{eq:fit}) is not reliable. Basically, under the conditions chosen, the low-frequency oscillation is undetectable in the gyrokinetic simulations in these two configurations. The usefulness of calculating the zonal-flow oscillation frequency with {\casdk} in the IS and OS configurations of LHD could be questioned in view that it is difficult to observe this oscillation in practice. However, this low frequency oscillation constitutes a natural mode of oscillation  that could manifest under particular conditions that enhance it and/or diminish the GAM oscillation. Oscillations in this frequency range could, in principle, be excited by some forcing mechanism, such as fast ions or electrons, and might be measured. With this idea in mind, the calculations shown in figure \ref{fig:zff-lhd-c3k} could be useful even in this kind of configurations.

\subsection{W7-X stellarator: high-mirror configuration}
\label{sec:w7xhmresults}

\begin{figure}
  \includegraphics[width=1.0\linewidth]{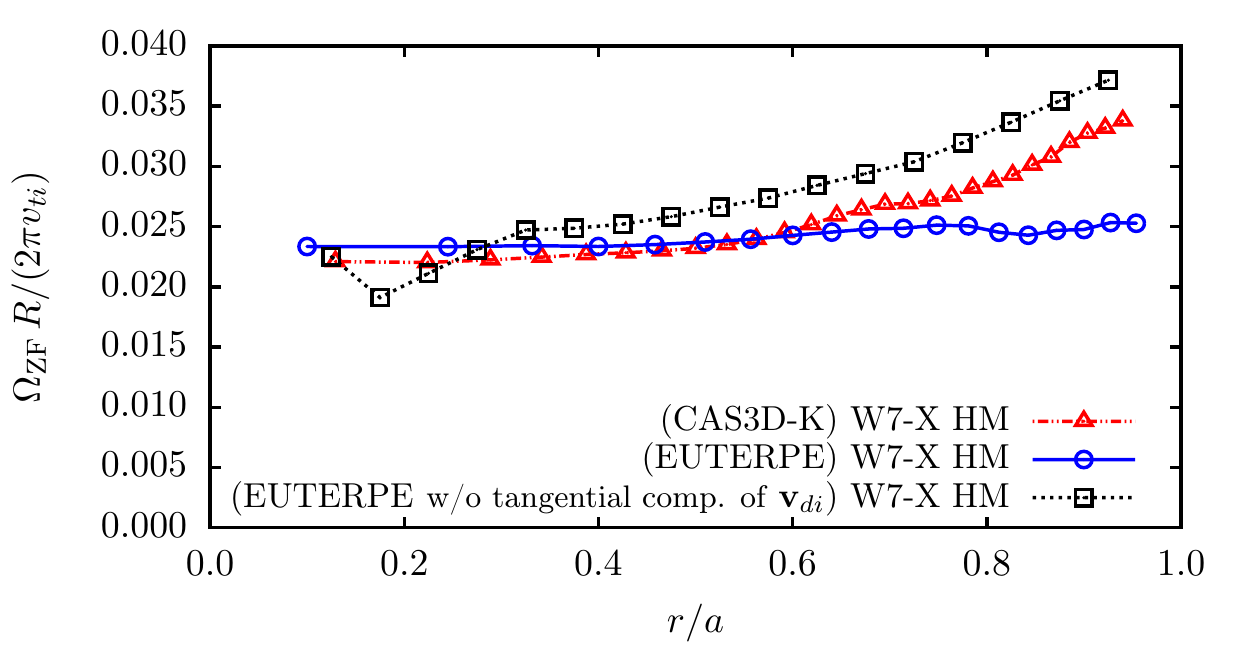}
  \caption{Zonal-flow frequency in the high-mirror configuration of the W7-X stellarator calculated with CAS3D-K and EUTERPE. The EUTERPE simulations employ $k_{\psi}=0.5 \pi$. As explained in the text, the EUTERPE simulations have been carried out in two settings: (a) including the full magnetic drift (blue curve); (b) retaining only the radial component of the magnetic drift (black curve).}
  \label{fig:zff-w7xhm}
\end{figure}

In figure~\ref{fig:zff-w7xhm} we show the values of $\Omega_{\rm ZF}$ for the high-mirror configuration of W7-X (W7-X HM) obtained with {\casdk} (red line) and {\eut} (blue line).  In the calculations presented here we use, as in the W7-X SD case, flat density and temperature profiles with $T_e=T_i=5$~keV. In the W7-X HM configuration the average magnetic field strength at the magnetic axis  is $B_0=2.35$~T. The $q$ profile is given in figure~\ref{fig:q-all}.  In the EUTERPE simulations we use an initial perturbation with $k_{\psi}=0.5\pi$, so that the normalized radial scale of the perturbation varies radially, with $\left\langle k_{\perp} \rho_{ti}\right\rangle_\psi < 0.032$. Note that in the W7-X high-mirror configuration, and for the indicated temperatures, the values of the frequency in physical units are on the order of  $\Omega_{\rm ZF}/(2\pi) \sim 3$~kHz, which are slightly larger than in the standard configuration.

The  {\casdk} and {\eut} calculations are in very good agreement for $r/a<0.6$, but not for $r/a>0.6$. In this case the difference between calculations of {\casdk} and {\eut} seem too large to be attributed to fitting errors or boundary condition effects in the gyrokinetic simulations. Let us try to understand the disagreement.

Note that in equation (\ref{eq:gk}) the component of the magnetic drift tangent to the flux surface has been neglected. This is formally correct as long as condition (\ref{eq:conditions_validity_gkeq}) is satisfied. We will see next, however, that in the W7-X HM configuration the neglected term contributes to the calculation of $\Omega_{\rm ZF}$. We cannot easily extend the analytical calculation of $\Omega_{\rm ZF}$ to include the effect of the tangential component of the magnetic drift, but we can carry out the {\eut} simulations removing this component, and therefore leaving only the radial component of the magnetic drift, which is a more faithful comparison to (\ref{eq:gk}).

The black curve in figure \ref{fig:zff-w7xhm} corresponds to the {\eut} simulation retaining only the radial component of the magnetic drift. In these simulations the damping of the low frequency oscillations  increases significantly with respect to the cases in which the full magnetic drift is kept. This makes the fit to a damped oscillation model like (\ref{eq:fit}) more difficult, but still possible. We can safely conclude that the role of the tangential magnetic drift in the W7-X HM configuration is needed to explain the radial dependence of $\Omega_{\rm ZF}$ for $r/a > 0.6$.

\begin{figure}
  \includegraphics[width=1.0\linewidth]{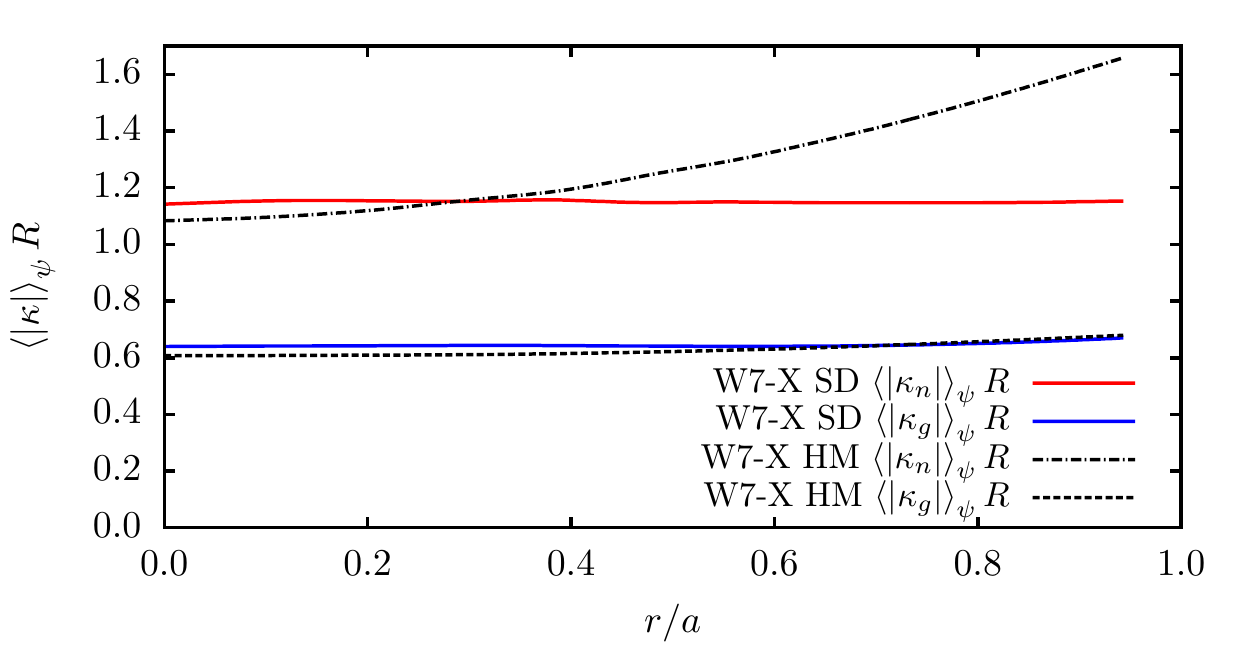}
  \caption{Flux-surface average of the absolute value of the geodesic and normal magnetic field line curvatures for the W7-X SD and W7-X HM configurations.}
  \label{fig:w7xstd-hmCurvs}
\end{figure}

The relevance of the tangential component of the magnetic drift in the W7-X HM configuration, by comparison to its irrelevance in the W7-X SD one, can be understood by looking at figure \ref{fig:w7xstd-hmCurvs}, where the flux-surface-averaged absolute value of the normal and geodesic components of the field line curvatures is shown for both configurations. While the geodesic component of the field line curvature is very similar in both cases, the normal component (related to the tangential component of the magnetic drift) is larger in the W7-X HM configuration than in the W7-X SD one. Actually, the difference is significant for $r/a \gtrsim 0.5$ and gets larger at outer positions (compare figure~\ref{fig:w7xstd-hmCurvs} with the black and blue curves in figure \ref{fig:zff-w7xhm}).

At this point, one might wonder why we do not do the same test for the TJ-II and LHD SD configurations. That is, one might think that perhaps it is possible to prove that the agreement between CAS3D-K and EUTERPE simulations cannot be improved for TJ-II and LHD SD because the tangential component of the magnetic drift also counts in these configurations. However, this check turns out to be not feasible.  In the simulations carried out neglecting the tangential component of the magnetic drift, the damping of the oscillations is remarkably higher with respect to the cases in which the full magnetic drift is retained. As we have already explained,  in  W7-X HM this effect is not enough to prevent a reasonable fit. On the contrary, in the rest of configurations studied (W7-X SD, TJ-II and LHD) the damping is so large when the tangential magnetic drift is removed that the fit is not reliable.

\section{Simulation details and computation time}
\label{sec:simulations}

One of the most important features of the semianalytical method herein proposed for the evaluation of the zonal flow oscillation frequency is that it is faster than calculating it by means of gyrokinetic simulations with {\eut} or {\gene}. We show in table~\ref{tab:cpuhts} the total CPU time required to obtain the frequency with each method. The times shown in the table for CAS3D-K and {\gene} correspond to the time required for calculations at just one radial position, while for {\eut}, as it is a global code, these values are given for the simulation of all radial positions at the same time. The values shown in table~\ref{tab:cpuhts} are determined by the resolution used in the different cases with {\casdk} and by the simulation time required to obtain a large enough number of oscillation cycles to make a fit in the case of {\eut} and {\gene}. In the {\eut} case, as they are global simulations, the minimum time required is determined by the most demanding radial position (the innermost radial locations, in general). These times represent the requirements for simulations or calculations converged with respect to the different numerical parameters and resolutions in each code. The numerical details in each case are listed below.

\begin{table}
  \centering
  \begin{tabular}{ r c l c}
    \hline 
    \hline
    & {\casdk} & {\eut} & {\gene} \\ 
    \hline 
        {\small LART (large ripple)} & $3-4$ & $\sim 2000$ & ---  \\
        {\small LART (small ripple)} & $2-3$ & $\sim 10000$  & --- \\
        {\small W7-X SD} & $2-3$ & $\sim 2000$  & $\sim 180$ \\
        {\small W7-X HM} & $2-3$ & $\sim 2000$  & --- \\
        {\small TJ-II SD} & $3-4$ & $\sim 1000$  & --- \\
        {\small LHD SD, OS \& IS} & $1-3 $ & $\sim 3000$  & --- \\ 
        \hline 
        \hline
  \end{tabular}
  \caption{CPU time (total core hours) required to obtain the zonal-flow frequency with CAS3D-K, EUTERPE and GENE in the magnetic configurations considered in the paper. Note that the CAS3D-K and GENE calculations are radially local and the values in the table correspond to the time needed for the calculation at a single radial position. EUTERPE is radially global and the times in the table correspond to the full radius calculation.}
  \label{tab:cpuhts}
\end{table}

In {\casdk}, the computation time to obtain converged results varies among devices and it depends on the phase-space resolution used in each case. The minimum resolution for the integration over the $\lambda$ coordinate in all the calculations is $n_\lambda=24$. The integration over the magnitude of the velocity is performed analytically. For the spatial integrals we use $n_\zeta=128$ trajectories and $n_\theta=512$ integration points per trajectory, in the case of passing particles. For trapped particles, these numbers are $n_\zeta=16$ trajectories per integration group and $n_\theta=512$ integration points per trajectory. The calculations in {\casdk} are radially local. Therefore, the values in the table corresponding to {\casdk} calculations are given per radial position and the ranges given in table~\ref{tab:cpuhts} account for the maximum and minimum CPU times among different radial positions at a given configuration.

\begin{figure}
  \includegraphics[width=1.\linewidth]{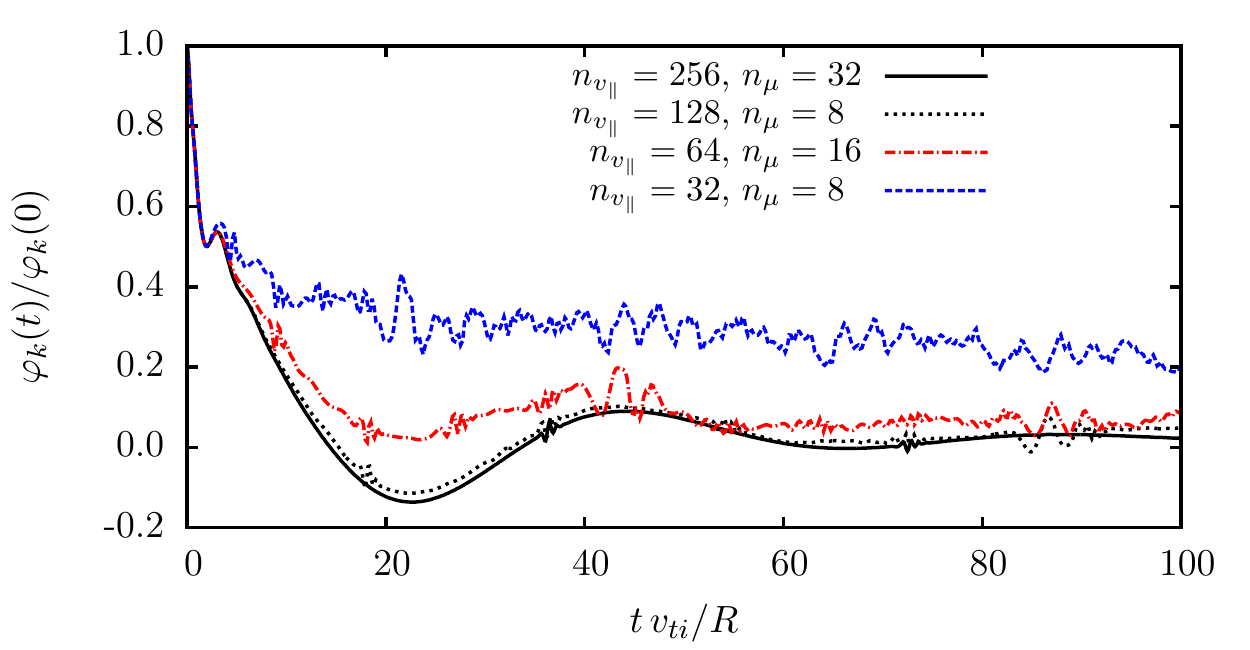}
  \caption{Time evolution of the electrostatic potential at $r/a=0.5$ in the W7-X SD configuration obtained with GENE for different resolutions in velocity space. In all cases, the radial scale of the perturbation is the same, $\left\langle k_{\perp} \rho_{ti}\right\rangle_\psi = 0.05$.}
  \label{fig:t-gene}
\end{figure}

The {\eut} simulations presented here are not extremely demanding from a computational point of view as only the zonal ($m=0$, $n=0$ Fourier component) and several smaller amplitude sidebands are resolved. The simulations were carried out with the following numerical parameters. In all the cases the resolution in poloidal and toroidal angles in PEST coordinates was $n_{\theta}=32$, $n_{\phi}=32$. The radial resolution was $n_s=24$ for the rippled tokamaks, TJ-II and LHD, while it was $n_s=32$ for the W7-X configurations. Simulations in W7-X were carried out with more radial resolution because several simulations with different radial scales of the perturbation, and maintaining the numerical parameters, were carried out for comparison. The number of markers used was $n_M=40\mathrm{M}-50\mathrm{M}$. 

In general, the {\eut} calculations require larger computational resources for devices with lower zonal-flow frequencies as more simulation time is required to resolve a number of oscillation cycles that is large enough to make a fit. For the rippled tokamaks discussed in Section~\ref{sec:Tokamakresults}, the required computational time with {\eut} increases by a factor of 5 for the case with smaller ripple with respect to that with the largest one because the oscillation frequency decreases by this factor, while all physical and numerical parameters in the simulation are the same. In the case of LHD configurations there is a large amplitude GAM oscillation, which makes difficult the fit of the potential time trace to a model like (\ref{eq:fit}) for the low frequency oscillation, which has much smaller amplitude. A longer simulation time was required as compared to other configurations to obtain a reasonable fit in the LHD standard configuration. Only in the standard configuration of LHD a reliable fit of the potential time traces was possible.

The {\gene} calculations were performed employing its full flux-surface version. Therefore, the calculations are 2D in the spatial coordinates $\{y,z\}$ and 2D in velocity space $\{v_\parallel, \mu\}$. Here, $y$ is the coordinate along the binormal direction, $z$ is the coordinate along the field line, $v_\parallel$ is the parallel velocity and $\mu$ is the magnetic moment coordinate. The results of the CPU time per radial position obtained with {\gene} in the W7-X SD configuration are given in table \ref{tab:cpuhts}. An analysis of convergence in velocity space resolution was carried out and the results are shown in figure \ref{fig:t-gene}. In that figure, several electrostatic potential time traces at $r/a=0.5$ are plotted, obtained from simulations with {\gene} in the W7-X SD configuration with $\left\langle k_{\perp} \rho_{ti}\right\rangle_\psi = 0.05$ and using different resolutions in velocity space. The CPU time given for {\gene} in table \ref{tab:cpuhts} corresponds to the simulation of figure \ref{fig:t-gene} with $n_{v_\parallel}=128$ and $n_\mu=8$. All the simulations were carried out with a very small value of hyperdiffusivity (two orders of magnitude smaller than typical values used in turbulence simulations), which is required to reduce the damping of the zonal flow oscillation and allow a reliable fit \cite{Pueschel2010}.

The calculations/simulations in this work were carried out in different supercomputers, with different capabilities, so that the computing time in the table~\ref{tab:cpuhts} has to be considered as indicative. The {\eut} simulations were carried out in two different supercomputers, EULER and MareNostrum III  \cite{MN3}. EULER is equipped with Intel Xeon 5450 quadcore processors at 3.0 GHz and Infiniband 4X DDR and Mare Nostrum III is equipped with Intel SandyBridge-EP processors at 2.6 GHz and Infiniband FDR10 interconnection. From 32 to 64 computing cores were used in the simulations. All the {\casdk} calculations shown in this work were run also in the EULER supercomputer. The {\gene} calculations were carried out in the Uranus supercomputer, which is equipped  with Intel Xeon E5-2630 processors at 2.4 GHz interconnected by Infiniband FDR. The computing time shown in the table is always the total CPU time (summed for all the computing cores used).

From the numbers shown in table~\ref{tab:cpuhts}, even corresponding to different computing facilities, it is clear that the calculations of the zonal-flow frequency with {\casdk} are faster than those employing the gyrokinetic codes {\eut} and {\gene}. This result strongly supports the usage of {\casdk} in this type of computations. 

\section{Discussion and conclusions}
\label{sec:conclusions}

In this work, we have proven the efficiency of a semianalytical method for calculating the zonal flow oscillation frequency in stellarators and rippled tokamaks. It is based on the numerical evaluation of expression (\ref{eq:zff}), which was first derived in \cite{Mishchenko2008} but had not been compared with the frequency obtained from direct gyrokinetic simulations so far.

We have extended  the code {\casdk} for the evaluation of expression (\ref{eq:zff}) in general rippled tokamak and stellarator configurations. In particular, we have implemented in {\casdk}, for both passing and trapped particles, the correct solution for $\delta_s$, the displacement from the initial flux surface at each point of the particle orbit. The accuracy of this semianalytical approach using the code {\casdk} was checked by comparing its results against the frequency obtained from gyrokinetic simulations with the global code {\eut} and the radially local code {\gene} in a wide range of  configurations. Specifically, we have calculated in a series of large aspect ratio tokamaks with different ripple values, as well as in the W7-X, TJ-II and LHD stellarators.

When using gyrokinetic codes, we obtain the zonal-flow frequency by fitting the  time trace of the normalized zonal electrostatic potential (or electric field) to a model function, given in (\ref{eq:fit}), including a damped oscillation and the decay to a residual value. The evolution of an initial zonal perturbation is initially dominated by a decaying GAM-like oscillation followed by a damped low frequency oscillation. Therefore, obtaining a precise value of the zonal-flow frequency from the fitting method is non-trivial, in general, and requires a case by case discussion.

The good agreement between the two methods (the semianalytical approach and the calculation of the frequency from gyrokinetic simulations) in real stellarator configurations supports the validity of the approximations in the derivation of (\ref{eq:zff}) and the accuracy of the semianalytical approach via the extension of the {\casdk} code. Only in the outer radial region of the high-mirror configuration of W7-X the accuracy of the semianalytical method is clearly insufficient. The reason is that, as we have explained, in this case, the tangential component of the magnetic drift, that is dropped in the analytical calculation leading to (\ref{eq:zff}), must be kept.

The advantage of using {\casdk} is that the computation time can be reduced up to two orders of magnitude with respect to the gyrokinetic calculations. This makes this method an option to be included in an stellarator optimization loop, in which {\casdk} could provide fast calculations of zonal flow relaxation properties (oscillation frequency and residual level \cite{Monreal2016}) to be used as figures of merit of stellarator configurations.

Finally, it is worth emphasizing that expression (\ref{eq:zff}) captures the influence of the magnetic geometry on the zonal flow oscillation frequency. However, as explained in subsection \ref{sec:longtimeevolutionZF}, this value of the frequency can be modified in the presence of a background radial electric field, $E_\psi$. The rigorous calculation of the corrections introduced by $E_\psi$ to expression (\ref{eq:zff}) are beyond the scope of this work.

\section*{Acknowledgments}
\label{sec:acknowledgements}

P.~M. thanks Per Helander and Tobias G\"orler for helpful discussions. The authors thank Antonio L\'opez-Fraguas for his help with the usage of {\small VMEC} and acknowledge the computer resources, technical expertise and assistance provided by the Barcelona Supercomputing Center (BSC) and the Computing Center of CIEMAT. {\gene} runs have been carried out in Uranus, a supercomputer cluster located at Universidad Carlos III de Madrid (Spain) funded jointly by EU FEDER funds and by the Spanish Government via the National Projects UNC313-4E-2361, ENE2009-12213-C03-03, ENE2012-33219 and ENE2012-31753.

This research has been funded in part by grants ENE2012-30832 and ENE2015-70142-P, Ministerio de Econom\'ia y Competitividad (Spain) and by an FPI-CIEMAT PhD fellowship. This work has been carried out within the framework of the EUROfusion Consortium and has received funding from the Euratom research and training programme 2014-2018 under grant agreement No 633053. The views and opinions expressed herein do not necessarily reflect those of the European Commission.

\section*{References}

\end{document}